\documentclass[a4paper,fleqn,usenatbib]{mnras}

\usepackage[T1]{fontenc}
\usepackage{ae,aecompl}
\usepackage{subfig}

\usepackage{graphicx}	
\usepackage{amsmath}	
\usepackage{amssymb}	

\title[Mass loss in tidally stripped systems]{Mass loss in tidally stripped systems; the energy-based truncation method}

\author[N. E. Drakos et al.]{
	Nicole E. Drakos$^{1,2,3} $\thanks{E-mail: ndrakos@ucsc.edu},
	James E. Taylor$^{2,3}$,
	and Andrew J. Benson$^{4}$
	\\
	$^{1}$Department of Astronomy and Astrophysics, University of California, Santa Cruz, 1156 High Street, Santa Cruz, CA 95064 USA \\	
	$^{2}$Department of Physics and Astronomy, University of Waterloo, 200 University Avenue West, Waterloo, ON N2L\,3G1, Canada \\
	$^{3}$Waterloo Centre for Astrophysics, University of Waterloo, 200 University Avenue West, Waterloo, ON N2L\,3G1, Canada \\
	$^{4}$Carnegie Observatories, 813 Santa Barbara Street, Pasadena, CA 91101, USA\\
}

\date{Accepted XXX. Received YYY; in original form ZZZ}

\pubyear{2020}

\begin{document}
	\label{firstpage}
	\pagerange{\pageref{firstpage}--\pageref{lastpage}}
	\maketitle
	
	\begin{abstract}
		The ability to accurately predict the evolution of tidally stripped haloes is important for understanding galaxy formation and testing the properties of dark matter. Most studies of substructure evolution make predictions {\color{black} based on} empirical models of tidal mass loss that are calibrated {\color{black} using} numerical simulations. This {\color{black} approach} can be accurate in the cases considered, but lacks generality and does not provide a physical understanding of the processes involved. Recently, we demonstrated that truncating NFW distribution functions sharply in energy results in density profiles that resemble those of tidally stripped systems, offering a path to constructing physically motivated models of tidal mass loss. In this work, {\color{black} we review calculations of mass loss based on energy truncation alone, and then consider what secondary effects may modulate mass loss beyond this. We find that a combination of dependence on additional orbital parameters and variations in individual particle energies over an orbit results in a less abrupt truncation in energy space as a subhalo loses mass. Combining the energy truncation approach with a simple prediction for the mass-loss rate, we construct a full model of mass loss that can accurately predict the evolution of a subhalo in terms of a single parameter $\eta_{\rm eff}$.}  This parameter can be fully determined from the initial orbital and halo properties, and does not require calibration with numerical simulations.
	\end{abstract}
	
	
	

\section{Introduction} \label{sec:Intro}

The standard theory of structure formation predicts that dark matter haloes grow through `hierarchical' merging, in which small structures merge to form progressively larger ones. Merging haloes evolve through dynamical friction, {\color{black} tidal} mass loss and tidal heating, and survive as self-bound structures {\color{black} (`subhaloes')} within galaxy, group, and cluster haloes. Isolated dark matter haloes have a nearly universal density profile (UDP), which is commonly described in terms of the Navarro-Frenk-White (NFW) profile \citep{navarro1996,navarro1997}:
\begin{equation}
\rho(r) = \dfrac{\rho_0 r_{\rm s}^3}{r(r+r_{\rm s})^2} \,\,\, ,
\end{equation}
where $\rho_0$ is a characteristic density and $r_{\rm s}$ is the scale radius, describing the point where the logarithmic slope is ${\rm d} \log \rho/ {\rm d} \log r = -2$. {\color{black} In subhaloes, this profile and the underlying distribution function are modified by heating and mass loss in more or less complex ways.}

Small scale substructure may be the best place to test the nature of dark matter. {\color{black} While numerical simulations of cold dark matter (CDM) can reproduce the clustering of galaxies and other probes of large scale structure very well, it is less clear how well they match observations on small scales (below $\sim$1 Mpc). The apparent discrepancies between theory and observations are likely due to baryonic feedback, but may also indicate departures from pure CDM \citep[e.g~][for a recent review]{bullock2017}}. For example, warm dark matter {\color{black} would} suppress the formation of low mass subhaloes \citep[e.g.][]{knebe2008} while self-interacting dark matter {\color{black} would produce less concentrated haloes with constant-density cores} \citep[e.g.][]{burkert2000,rocha2013}. {\color{black} To constrain both these possibilities, it is essential to understand how the smallest, densest subhaloes evolve within the tidal field of larger haloes.}

{\color{black} Another} promising test of dark matter properties is the annihilation signal predicted for some candidates, {\color{black} notably supersymmetric weakly-interacting massive particles (WIMPs).} Since the annihilation rate {\color{black} is} proportional to the square of the local dark matter density at any point, it is sensitive to the inhomogeneity of the particle distribution. Substructure in a halo will contribute greatly to the overall annihilation signal, and establishing robust constraints on the cross-section for annihilation requires accurate predictions of subhalo evolution \citep[e.g.][]{lacroix2018, hiroshima2018, stref2019,hutten2019,delos2019d}. An issue of particular importance is the long-term survival of the very smallest structures over many {\color{black} orbital periods, as these can dominate the overall annihilation rate} \citep[e.g.][]{okoli2018}.

Though most of the understanding of dark matter haloes comes from {\color{black} self-consistent cosmological simulations, idealized simulations of isolated systems are also used} to understand the detailed physical processes involved in halo mergers. Tidal stripping in minor mergers, for instance, is frequently studied using $N$-body simulations of a satellite halo evolving within a static host halo \citep[e.g.][]{hayashi2003, kazantzidis2004, boylankolchin2007, kampakoglou2007, penarrubia2008a, penarrubia2008b, penarrubia2009,choi2009, penarrubia2010, drakos2017, ogiya2019,delos2019b}. These idealized simulations provide a way to test the physical processes that govern subhalo evolution and determine the long-term fate of subhaloes as they lose mass, but they normally require interpretation with analytic models in order to {\color{black} extrapolate} the results {\color{black} to the full range of situations relevant in calculations of annihilation rates, substructure lensing and semi-analytic models of galaxy evolution}.

{\color{black} From this body of work, a number of empirical descriptions of subhalo evolution have been derived \citep[e.g.][]{hayashi2003,penarrubia2010,green2019,delos2019b}, as well as a few attempts} {\color{black} to describe the evolution from first principles \citep[e.g.][]{kazantzidis2004,drakos2017}. The two approaches are complimentary; the former are often easier to implement, and can be more accurate in the specific cases simulated, while the latter give insight into the physical processes involved, and can be applied to a wider range of scenarios than those simulated (e.g. different host and subhalo profile models and a larger range of orbital parameters).}
 Unfortunately, there is no rigorously correct analytic model for mass loss, as none of the simplifying assumptions required to derive one hold in a time varying, non-axisymmetric potential. Nonetheless, behaviour in limiting cases like the Jacobi model or the impulse approximation can provide guidance. 
  
 Overall,  given a clear definition or way of distinguishing bound and unbound particles (which itself is problematic), {\color{black} an analytic model of mass loss needs to explain and predict three things: (1) which particles are lost, or in what order particles are lost from a system (what criterion and which particle properties are relevant) (2) how the structure responds to mass loss and how the properties of individual particles change as mass is lost, and (3) the overall mass-loss rate. In \cite{drakos2017}, hereafter Paper I, we addressed the first two points, proposing that particles are lost based on their binding energy, but did not estimate the mass-loss rate.} The main goal of this paper is to provide an estimate for the mass-loss rate, and show that it matches simulations without adjustment of free parameters. Before we do so, we will first examine the assumptions and limitations of the energy-truncation or `lowering' approach. Putting these pieces together will produce a complete (albeit simplified) picture of tidal mass loss. 

Classical approaches to modelling tidal mass loss generally truncate haloes spatially, removing {\color{black} some or all} material outside some tidal radius \citep[e.g.][]{taylor2001a, benson2002, penarrubia2004, just2005, jiang2016}, though more detailed models of subhalo evolution also take into account individual particle energies and angular momenta \citep[e.g.][]{kampakoglou2007}. However, as first pointed out by \cite{choi2009}, tidal stripping may be better described as a truncation process in energy space. Pursuing this finding, in Paper I, we {\color{black} described an approach to modelling mass loss based} on an energy truncation of a halo's distribution function (DF). In this model, mass loss occurs in discrete steps once per orbit (at pericentre), and produces a progressive truncation in energy of the DF at each step. The system is assumed to {\color{black} relax back to an} isotropic and spherical {\color{black} form at apocentre}; it responds to mass loss by simply `lowering' the DF by a constant shift in particle energies. Aside from this overall shift, particle energies are conserved from one orbit to the next, such that the next phase of stripping at pericentre can be calculated from the particle properties at the preceding apocentre.
 
When applying the energy-truncation model to isolated simulations of NFW subhaloes, we found it matches the stripped density profile and even the phase-space distribution quite well. Additionally, this model is relatively universal (i.e. that it applies to any density profile or DF), as we will explore further in a forthcoming paper, Paper III. We showed that while the model works surprisingly well to predict the profile evolution for cases of slow mass loss, it is less accurate in cases of rapid mass loss. In this work, we explore the energy-based approach further, to gain a deeper understanding of its limitations. We show in particular that particle energies calculated at successive apocentres are not completely conserved, and that this adds scatter to the truncation in terms of the original particle energies. Given the relative success of the energy truncation approach, an estimate of the mass-loss rate is still needed to predict the evolution of the subhalo. In this paper we review various estimates of the mass-loss rate and show that a simple, non-parameteric estimate, when combined with energy truncation, does an excellent job of predicting the mass-loss rates seen in our simulations.

The structure of this paper is as follows: in Section~\ref{sec:model} {\color{black} we review the approach to modelling mass loss introduced in Paper I, by truncation of the distribution function in energy space, and} in Section~\ref{sec:NFWSims} we briefly summarize the simulations from Paper I. In Section~\ref{sec:compare}, we compare {\color{black} radial profiles predicted by energy truncation to those predicted by other mass-loss models in the literature.} In Section~\ref{sec:lim} we explore the {\color{black} validity of our approximations and limitations of the energy-based approach}, to gain a thorough understanding of the main determinants of subhalo profile evolution. Then, in Section~\ref{sec:massloss}  we {\color{black} combine the idea of energy truncationwith a simple estimate of the mass loss rate to produce a full predictive model of mass loss that works well} with no further calibration or adjustment. Finally, we summarize the findings of this work in Section~\ref{sec:conc}.

\section{Review of energy-truncation model} \label{sec:model}

In Paper I, we outlined a method for truncating a spherically symmetric system in energy space, which we will summarize here. First, we provide a brief review of the DF and its properties; a more detailed description can be found in \cite{binney}.

\subsection{Distribution function}

The DF, $f(r,v)$, describes the mass per phase-space volume element ${\rm d}r^3{\rm d}v^3$. For spherically symmetric, isotropic systems, the DF {\color{black} depends only on a single variable, the `relative energy'} $\mathcal{E}= \Psi(r) - v^2/2$; i.e. $f(r,v) = f(\mathcal{E})$. The relative potential energy {\color{black} used here}, $\Psi(r)$ is defined as $\Psi =-( \Phi + \Phi_0)$, where $\Phi_0$ is usually defined at the outer boundary of the system, such that $f>0$ when $\mathcal{E}>0$, and $f=0$ otherwise. {\color{black} Given the sign convention, the relative energy is equivalent to the (positive) binding energy required to eject a particle from the system to infinity; we will use relative energy and binding energy interchangeably in what follows.} The DF is related to the density profile, $\rho(r)$, as follows:
\begin{equation} \label{eq:rho_DF}
\rho(r) = 4 \pi \int_0^{\Psi(r)} f(\mathcal{E}) \sqrt{2(\Psi(r)-\mathcal{E})} \rm d \mathcal{E} \,\,\,.
\end{equation}

This relationship can be inverted, and the DF expressed in terms of the density profile by using Eddington's inversion method \citep{eddington1916}:
\begin{equation}
f(\mathcal{E}) = \dfrac{1}{\sqrt{8}\pi^2} \left[ \int_0^\mathcal{E} 
\dfrac{1}{\sqrt{\mathcal{E} - \Psi}} \dfrac{{\rm d}^2 \rho}{{\rm d}\Psi^2} \rm d \Psi + \dfrac{1}{\sqrt{\mathcal{E}}} \left( \dfrac{{\rm d}\rho}{{\rm d}\Psi}\right)_{\Psi=0}
\right] \,\,\, .
\end{equation}

\subsection{The model} \label{sec:ourmodel}

The model presented in Paper I is created by lowering the original DF of the subhalo and then recovering the modified density profile, analogously to how the King Model is created by lowering the DF of an isothermal sphere \citep{king1966}. This method was originally proposed in \cite{widrow2005} as a method to truncate NFW profiles for use as initial conditions (ICs) in isolated simulations. In Paper I, we showed that it also works well as a description for tidally stripped haloes. 

First, given the DF of the original system $f_0(\mathcal{E})$, the new DF is expressed as:
\begin{equation} \label{eq:fE}
f(\mathcal{E}) = 
\begin{cases}
f_0(\mathcal{E}+\mathcal{E}_T)-f_0(\mathcal{E}_T)& \mathcal{E} \ge 0\\
0& \mathcal{E} \le 0  \,\,\, ,
\end{cases}
\end{equation}
where $\mathcal{E}_T$ is the truncation or tidal energy. Then, the relative potential of the truncated system can be found by solving Poisson's equation using standard techniques for solving second-order ODEs:
\begin{equation} \label{eq:Psi}
\begin{aligned}
\dfrac{{\rm d}^2{\Psi}}{{\rm d}r^2} + \dfrac{2}{r} \dfrac{{\rm d}\Psi}{{\rm d}r} &=-16 \pi^2 G \int_0^{\Psi(r)} f(\mathcal{E}) \sqrt{2(\Psi(r)-\mathcal{E})} {\rm d} \mathcal{E}\\
\\
\Psi(0) &= \Psi_0(0) - \mathcal{E}_T \\
\dfrac{{\rm d}\Psi(0)}{{\rm d} r} &= 0 \,\,\, ,
\end{aligned} 
\end{equation}

Where $\Psi_0$ is the relative potential of the original, un-truncated system. Once $\Psi$ has been calculated, the truncation radius, $r_t$, is given by $\Psi(r_t)=0$, and the density profile can be found from Equation~\eqref{eq:rho_DF}.

\section{Simulations} \label{sec:NFWSims}

In this work, we use the idealized simulations from Paper I, which we will briefly summarize here. These simulations were performed using the $N$-body code \textsc{gadget-2} \citep{gadget2}, which was modified to contain a fixed, {\color{black}infinitely extended}, background potential corresponding to a host halo with an NFW profile {\color{black} of concentration $c_{\rm host} = 10$}. We use the mass and scale radius of the satellite halo ($M_{\rm sat}$ and $r_{\rm s}$) as the mass and distance units, {\color{black} and a softening length of $\epsilon = 0.01\,r_{\rm s}$.} Time is given in units $t_{\rm unit} =\sqrt{r_{\rm s}^3/GM_{\rm sat}}$, and velocity in units of $v_{\rm unit} = \sqrt{GM_{\rm sat}/r_{\rm s}}$. The host and satellite haloes were assumed to have the same initial density within their outer, or `virial' radii, as would be the case for a merger between two cosmological haloes at a fixed redshift; then, the virial radius of the main halo scales as $(M_{\rm host}/M_{\rm sat})^{1/3}$. 

The satellite halo was modeled as an NFW halo with concentration $c=10$. Initial conditions were generated using the publicly available code \textsc{icicle} \citep{drakos2017}, which assigns positions and velocities to each particle by sampling the mass profile and DF of a specified profile. Since the mass of NFW profiles diverges with increasing radius, particles were generated within a specified radius, $r_{\rm cut}=c r_{\rm s}$, and unbound particles were iteratively removed. The resulting satellite is a truncated NFW profile with $N=1286991$ particles, with a maximum radius of $r_{\rm cut}=10\, r_{\rm s}$, where $r_{\rm s}$ is the scale radius of the satellite. As shown in Paper I, this method results in a halo that looks remarkably similar to one created using the energy-truncation procedure outlined in Section~\ref{sec:model} (for this specific set of ICs, it is similar to a lowered NFW profile with a tidal energy of $\mathcal{E}_T \approx 0.28$).

Overall we considered four different host/satellite mass ratios, $M_{\rm host}/M_{\rm sat}=300,\,100,\,50$ and $10$. We also considered various orbital energies and angular momenta; these simulations are summarized in Table~\ref{tab:sims}.

\begin{table*}
	\caption{\label{tab:sims} Summary of simulation parameters. Columns give (1) the simulation number (2) the mass ratio between the host and satellite halo (3) the virial radius of the host (4) the apocentric distance (5) the pericentric distance (6) the tangential velocity at apocentre (7) the (radial) orbital period,  (8) the circularity of the orbit,  {\color{black}$\epsilon_c= L/L_{\rm max}$ (as defined in Section~\ref{sec:trunc})}, (9) the relative energy, $\eta_{\rm rel}$, (defined as the energy divided by the energy of a circular orbit at the virial radius) and  (10) the radius of a circular orbit with the same energy divided by the virial radius. {\color{black}Simulations 3 and 4 correspond to the Fast and Slow Simulations, respectively.}}
	\begin{tabular}{ c c c c c c c c c c}
		\hline
		Simulation & $M_{\rm host}/M_{\rm sat}$& $R_{\rm vir}/r_{\rm s}$ & $r_a/r_{\rm s}$ & $r_p/r_{\rm s}$ &$v_a/v_{\rm unit}$& $t_{\rm orb}/t_{\rm unit}$&  $\epsilon_c$ &  $\eta_{\rm rel}$ &  $R_c/R_{\rm vir}$ \\ \hline
		1 & 100 & 46.4 & 100 & 10 & 0.34 & 206.8 & 0.42 & 0.85 & 1.26 \\ 
		2 & 100 & 46.4 & 100 & 50 & 0.90 & 299.4 & 0.92 & 0.71 & 1.63\\
		3  {\color{black}(Fast Sim)} & 300 & 66.9 & 100 & 10 & 0.51 & 129.7  & 0.40 & 1.09 &0.88 \\
		4 {\color{black} (Slow Sim)} & 300 & 66.9 & 100 & 50 & 1.42 & 185.4 & 0.92 & 0.92 & 1.13 \\
		5 & 100 & 46.4 & 500 & 50 & 0.23 & 1778.5 &  0.47  & 0.24 & 6.14\\
		6 & 300 & 66.9 & 25 & 10 & 1.50 & 31.48& 0.82  & 2.29 & 0.26 \\
		7 & 50 & 36.8 & 80 & 5 & 0.19 & 201.6& 0.30  & 0.86  & 1.24 \\
		8 & 50 & 36.8& 90 & 15 & 0.37 & 259.7 & 0.58  & 0.76 & 1.48 \\
		9 & 10 & 21.5 & 40 & 10 & 0.30 & 196.8 & 0.71  & 0.88 & 1.19 \\
		10 & 10 & 21.5 & 25 & 10 & 0.42 & 123.2 & 0.85  & 1.14 & 0.82 \\
		\hline
	\end{tabular}	
\end{table*}

The bound satellite was defined as in Paper I. First, the centre of the satellite was found by calculating the centre of mass in increasingly smaller spheres, as originally described in \cite{tormen1997}. {\color{black}The sphere was initially given a radius of $r = r_{\rm cut}$ and then} decreased by $0.9\,r/r_{\rm s}$ at each iteration, until there were fewer than 100 particles in the sphere. The  velocity frame was calculated as the average particle velocity within the original $r_{\rm cut}$. Finally, the energy of each particle was then calculated in this frame (assuming a {\color{black}spherical potential}{\footnote{\color{black}We found that there was surprisingly little difference between assuming a spherical potential and calculating the full potential---see Appendix~\ref{sec:potential}}
{\color{black}and using only the bound particles in the calculation}), and unbound particles were iteratively removed until convergence.

In Paper I we found that our model works well for haloes that lose mass slowly, and less well for subhaloes that {\color{black} lose mass rapidly}, though the cause of this was unclear. In the following two sections, we will use Simulations 3 and 4 from Paper I, which are representative of simulations which lose mass quickly and slowly, and refer to them as the `Fast Simulation' and `Slow Simulation', respectively.

\section{Model comparisons} \label{sec:compare}

Here we compare our model's predictions for the density profile evolution in tidally stripped haloes to empirical fits in the literature \citep{hayashi2003,penarrubia2010,green2019}; detailed descriptions of each of these models are given in Appendix~\ref{sec:models}. 

\cite{hayashi2003} were the first to examine how tidally stripped haloes evolve using idealized simulations. They proposed a simple empirical model for tidally stripped NFW profiles, that could be described by one parameter, the bound mass fraction. A slightly more complicated empirical model was proposed by  \cite{penarrubia2010}; again, it depends only on the bound mass fraction. While the model proposed in \cite{penarrubia2010} is more complicated than that in \cite{hayashi2003}, it is also more general, and can be used for any profile of the form:
\begin{equation} 
	\rho(r) = \dfrac{\rho_0}{(r/{r_{\rm s}})^{\gamma}[1 +(r/r_{\rm s})^{\alpha}]^{(\beta - \gamma)/\alpha}}\,\,\,.
\end{equation}

More recently, \cite{green2019}, provided an updated version of the fits given in \cite{hayashi2003}; in this work they used the DASH smulations \citep{ogiya2019}, which are a suite of idealized minor merger simulations between NFW haloes. The advantages of these simulations over those used in \cite{hayashi2003} are that they have been carefully calibrated against numerical effects, they have initial conditions constructed from the full theoretical DF, and they consider different satellite concentrations. While this work produced much more accurate predictions of subhalo evolution, it was again limited to the case of NFW profiles.

Like the models proposed in \cite{hayashi2003}, \cite{penarrubia2010} and \cite{green2019}, our model is also dependent on a single parameter (expressed as the tidal energy, $\mathcal{E}_T$, but which can equivalently be expressed in terms of the bound mass). However, it has the advantages that the main parameter has an obvious (and testable) physical interpretation ($\mathcal{E}_T$ being the maximum {\color{black} binding energy, $\mathcal{E}$,} of any bound particle in the frame of the satellite {\color{black} before stripping}), and that it can potentially be applied to any collisionless system, regardless of density profile or distribution function---the universality of the model will be explored further in Paper III.

To compare how well our model describes the density profile of the subhalo remnant compared to other models, we fit the bound subhalo remnant after one orbit, at apocentre. Apocentre was chosen since the subhalo should be roughly in equilibrium at this point; similar results can be found in subsequent orbits (see Paper I). The tidal energy, {\color{black}$\mathcal{E}_T$}, for our model was calculated as in Paper I, by requiring that the total mass of the bound halo was equal to that of the model. The other three models were calculated as described in Appendix~\ref{sec:models}; in all cases, the fits were dependent on a single parameter, the bound mass.

Fig.~\ref{fig:NFW_compare} shows how the density, enclosed mass and circular velocity profiles of tidally stripped haloes in the simulations compare to our model, as well as to the models presented in \cite{hayashi2003} and \cite{penarrubia2010}. Residuals of the model fits are shown for the density, mass and circular velocity profiles, calculated as $(y_{\rm model}-y_{\rm simulation})/y_{\rm simulation}$. As in Paper I, our model is a better approximation to the remnant from the Slow Simulation then the one from the Fast Simulation, though in both cases it over-predicts the mass and density in the centre of the halo. Conversely, the model from \cite{hayashi2003} tends to under-predict the central mass and density of the central halo; this is likely because their simulations used the local Maxwellian approximation to generate their ICs, which can lead to artificial relaxation in the centre of the subhalo \citep{kazantzidis2004}. Overall, it appears that the fits from \cite{green2019} perform slightly better than all other models, but our model is at least comparable to, if not more accurate than the models presented in \cite{hayashi2003} and \cite{penarrubia2010}.

\begin{figure*}
	\includegraphics[]{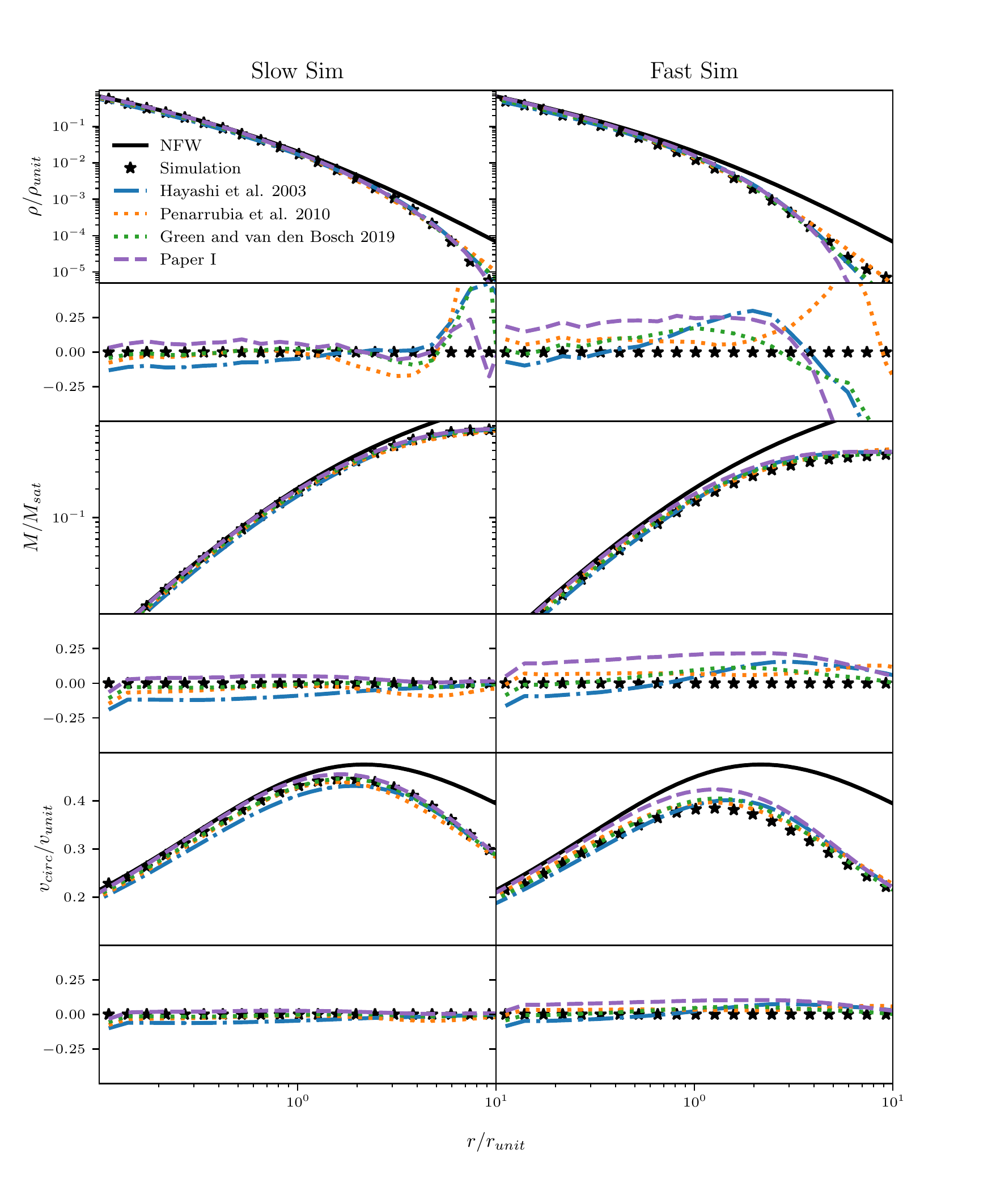}
	\caption{Comparison of {\color{black} different model descriptions of the} density, enclosed mass and circular velocity profiles of the bound satellite remnant after one orbit in the Slow (left) and Fast Simulation (right). Simulation results are shown with points, and the models are shown with lines. The thick dashed curve shows the original, untruncated NFW profile. The residuals in the fits are calculated as ($y_{\rm model}/y_{\rm simulation}-1$), where $y$ is the density, mass or circular velocity profile. Our model from Paper I is described in Section~\protect\ref{sec:model}, and the other models in Appendix~\protect\ref{sec:models}. {\color{black}In all cases, the models are dependent on one parameter, the bound mass fraction, which we calculate as the self-bound mass in the simulation, as described in the text. }
	}
	\label{fig:NFW_compare}
\end{figure*}

We note that since all of these methods are dependent on the bound mass, they are sensitive to the method used for defining bound particles. Also, bound mass calculations include mass that is only temporarily bound, but is leaving the system \citep[e.g.][]{penarrubia2009}. Most of this temporarily-bound mass is at larger radii, which may explain why the mass loss models all fail to match the bound mass profiles from the simulations beyond 4-5 $r_{\rm unit}$ in Fig.~\ref{fig:NFW_compare}. A more accurate normalization procedure for our model might be to fit the radius enclosing some fraction of the mass, or to fit the peak circular velocity; rather than take either of these approaches, we will later use the mass loss model in Section~\ref{sec:massloss} to predict $\mathcal{E}_T$ theoretically.

\section{Our model assumptions} \label{sec:lim}

The truncation method described in Section~\ref{sec:model} determines which particles will remain bound, and which will be ejected from the system, given a tidal truncation energy $\mathcal{E}_T$. As shown in the previous section, this approach predicts the density (or equivalently the mass or the circular velocity) profiles of the remnants quite well, but appears to be more accurate when mass loss is slower. Our goal in this section is to understand why this is the case, focusing on the Fast and Slow Simulations described in  Section~\ref{sec:NFWSims}. 

Suppose we consider a simple model where mass loss is primarily an energy-based process, as proposed by \cite{choi2009}. Three main simplifying assumptions are:

\begin{enumerate}
\item Mass loss occurs episodically around the pericentre of the orbit; after each episode the system returns to equilibrium, with the modified spherically symmetric, isotropic distribution function predicted by the energy truncation method.

\item Individual particles are lost or retained based solely on their energy with respect to the rest of the satellite.

\item At least over a single orbit, particle properties change only slowly; there is relatively little mixing of the energy stratification within the system, and thus which particles are lost during the orbit can be predicted from their energies at apocentre at the start of the orbit.
\end{enumerate}

We will test these assumptions below, starting with (i), considering how close the system is to equilibrium, and how its true shape and (an)isotropy affect mass loss at apocentre. Then we will test how cleanly mass loss depends on energy at the previous apocentre (ii). Finally, we will  consider assumption (iii), which we will show is the most problematic.


\subsection{Equilibrium, sphericity, and isotropy} \label{sec:equil}

\textbf{{\color{black}Equilibrium:}} One assumption of our model is that the subhalo is in approximate equilibrium at each apocentric passage. To test this, we identify the bound subhalo remnant at $t=2\,t_{\rm orb}$, and evolve it in isolation using \textsc{gadget-2} for an additional $t= 2\,t_{\rm orb}$. We compare the initial subhalo to the final subhalo in Fig.~\ref{fig:Equil_dens}. For both the Fast and Slow Simulations there is very little evolution, except at large radii. Since differences between the Fast Simulation and our model can be seen at all radii, this suggests that the assumption of equilibrium is is not the source of the differences between the two simulations.

\begin{figure}
	\includegraphics[clip=true,width = \columnwidth]{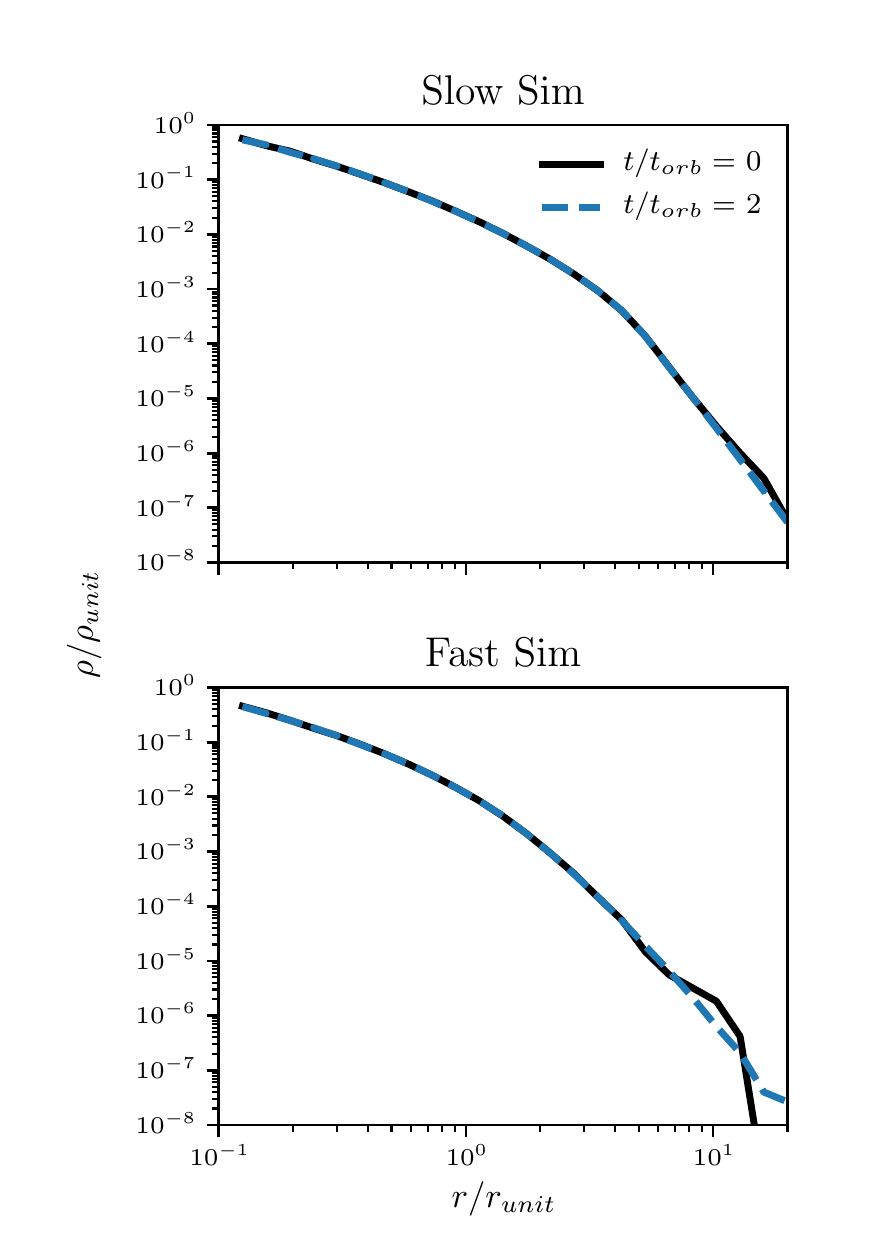}.
	\caption{{\color{black}Test of whether subhalo is in equilibrium at apocentre; the plot shows the evolution of the} subhalo density profile when removed from the fixed background potential. The initial haloes (solid black lines) are from the subhalo after two orbits in a fixed potential. The final profiles (dashed purple lines) are after the haloes are allowed to evolve for an additional $t=2t_{\rm orb}$ in isolation.}
	\label{fig:Equil_dens}
\end{figure}

\textbf{{\color{black}Sphericity:}}} Another assumption of our model is that the haloes are spherically symmetric at apocentre. To examine this point, the shape of the subhaloes were calculated as in \cite{dubinski1991,drakos2019a};  beginning with axis ratios $s=b/a=1$ and $q=c/a=1$ (where $a>b>c$ are the principal axes sizes), the dimensionless inertia tensor was calculated as $I_{ij} = \sum x_i x_j /d^2$, where $d = x_i^2 + (x_j/s)^2 + (x_k/q)^2$ is the ellipsoidal coordinate. The coordinates of each particle were rotated using the eigenvectors of {\color{black}$I$}, and the principal axis ratios $s$ and $q$ were recalculated. This process was repeated until convergence, which was defined as when $s=b/a$ and $q=c/a$ both had a percentage change of less than $10^{-3}$. 

\begin{figure}
	\includegraphics[clip=true,width = \columnwidth]{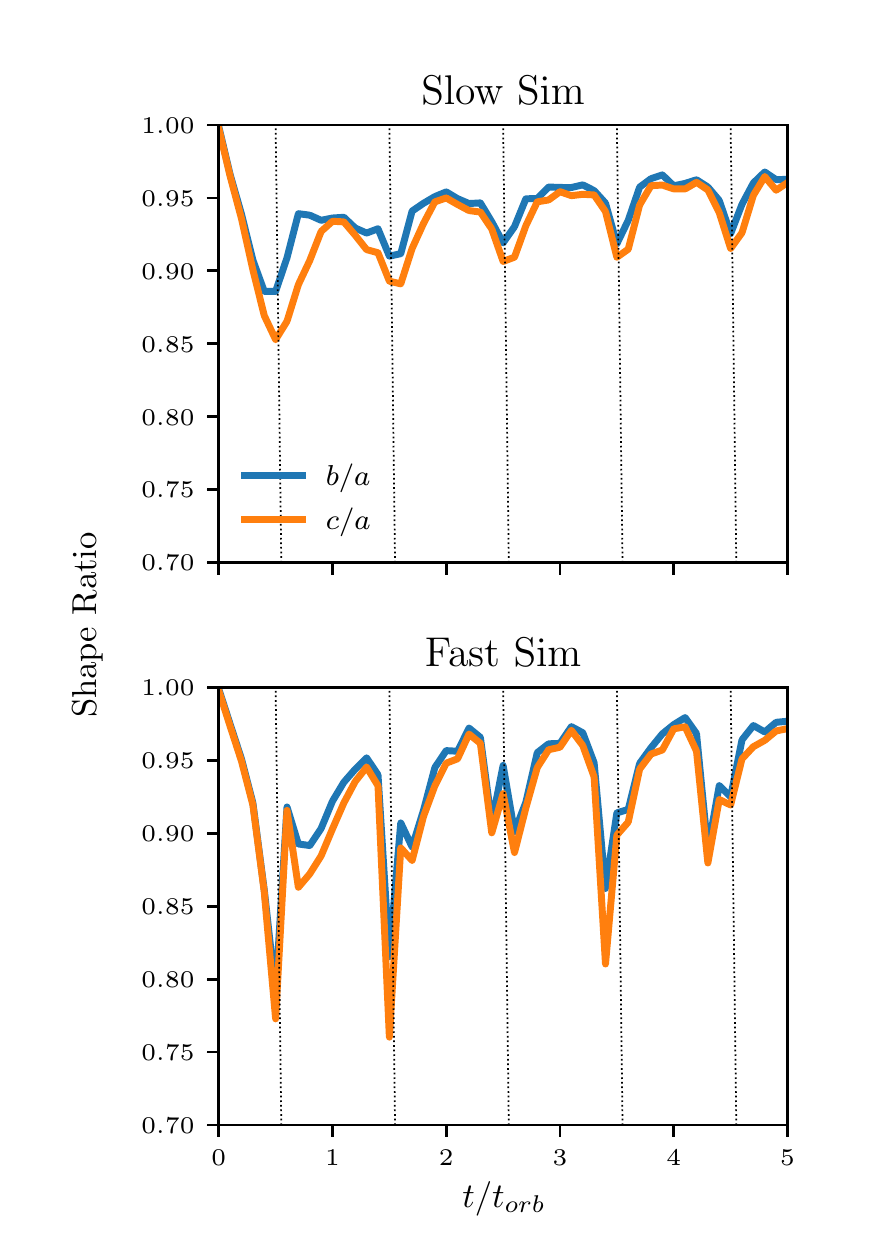}.
	\caption{Subhalo shape ratios as a function of time. Dotted lines indicate times of pericentric passage. {\color{black} In both cases, the subhaloes are roughly spherical at apocentre, though they become more elongated at pericentre}.}
	\label{fig:Shape}
\end{figure}

Fig.~\ref{fig:Shape} shows how the shape of the subhaloes evolves with time. In both the Fast and Slow Simulations the remnants are roughly spherical at apocentre. At pericentre, the shapes become more prolate ($a>b \approx c$), particularly for the Fast Simulation.  Since most mass loss occurs around this point, the difference between the Fast and Slow Simulations could reflect to the subhalo shape at pericentre. For instance, \cite{moore2004} showed that prolate haloes can lose mass at a rate several times higher than an isotropic spherical halo with the same density profile, and conclude {\color{black} that this is because the particles supporting the shape of the system are on radial orbits that are more vulnerable to being stripped. Overall, we conclude that the assumption that the remnant is spherical at apocentre is valid, but examine the effects of shape on stripping further in Appendix~\ref{sec:shape}.}

\textbf{{\color{black}Isotropy:}}
In addition to assuming the remnant is spherical at apocentre, our model also assumes they are isotropic, and thus the DF is only a function of the energy, $\mathcal{E}$.  {\color{black} The ICs ($t/t_{\rm orb}=0$) are isotropic by construction.  In Paper I, we examined the evolution of the anisotropy parameter, $\beta=1 - (\sigma_\theta^2 + \sigma_\phi^2)/2\sigma_r^2$ for the Fast and Slow Simulations (c.f. Figure~10). We showed that as time progresses, the remnants become more anisotropic, particularly at large radii.  {\color{black} Since, unlike the anisotropy, the ability of the model to describe the simulations does not depend on the number of pericentric passages, it} seems unlikely that this is the reason for the difference between the two simulations.

\subsection{Particle removal criteria}\label{sec:trunc}

In our model, particles are removed based strictly on their energy. It is known, however, that the other orbital parameters also play a role; particles on prograde orbits are more easily stripped than those on retrograde orbits, while particles on radial orbits are more easily stripped than those on circular ones \citep[e.g.][]{read2006a}. Since the energy-truncation model assumes that particles are removed based solely on energy, it ignores these complications. {\color{black} In this section we explore to what extent the truncation should depend on secondary particle properties, such as angular momentum and inclination angle.}

To explore the effect of particle angular momentum, as in \cite{choi2009}, we calculate initial the relative energy, $\mathcal{E}$, and circularity, $\epsilon_c$, for each particle as $\mathcal{E} = \Psi(r) - v^2/2$ and $\epsilon_c = L/L_{\rm max}$. To calculate $L_{\rm max}$,  we first determined the radius of a circular orbit with the same energy, $r_\mathcal{E}$ from $\mathcal{E} = \Psi(r_\mathcal{E}) -v_c(r_\mathcal{E})^2/s$, where $v_c$ is the circular velocity, and then  $L_{\rm max} = \sqrt{G M(r_\mathcal{E})r_\mathcal{E}}$. 

 {\color{black}The top panel of Fig.~\ref{fig:EPhasePlots}} shows the fraction of particles that remain bound to the remnant in $\epsilon_c$--$\mathcal{E}$ space, according to their original energy and angular momentum. From Fig.~\ref{fig:EPhasePlots}, it seems that there is a very slight angular momentum dependence in which particles are stripped, as in \cite{choi2009}. {\color{black} In the bottom panel of Fig.~\ref{fig:EPhasePlots} } we consider the bound fraction in inclination angle--energy space. The inclination angle was calculated as $\theta = \arccos L_z/|\vec{L}|$. By convention, $\theta<\pi/2$ corresponds to prograde orbits, while $\theta>\pi/2$ corresponds to retrograde orbits. We do see a dependence on inclination angle, but only in the Fast Simulation, and most notably after the first orbit. This suggests that prograde orbits are especially vulnerable to being stripped in cases of fast mass loss. This is perhaps not surprising, given the stronger tidal forces, though we suspect it also has to do with the 3D shape changes of the satellite at pericentre. {\color{black} Interestingly, the sharpness of the truncation seems to differ in the two cases, but cannot be fully explained by the angular momentum or inclinations of the particles.}

\begin{figure*}
	\centering
	\subfloat{{\includegraphics{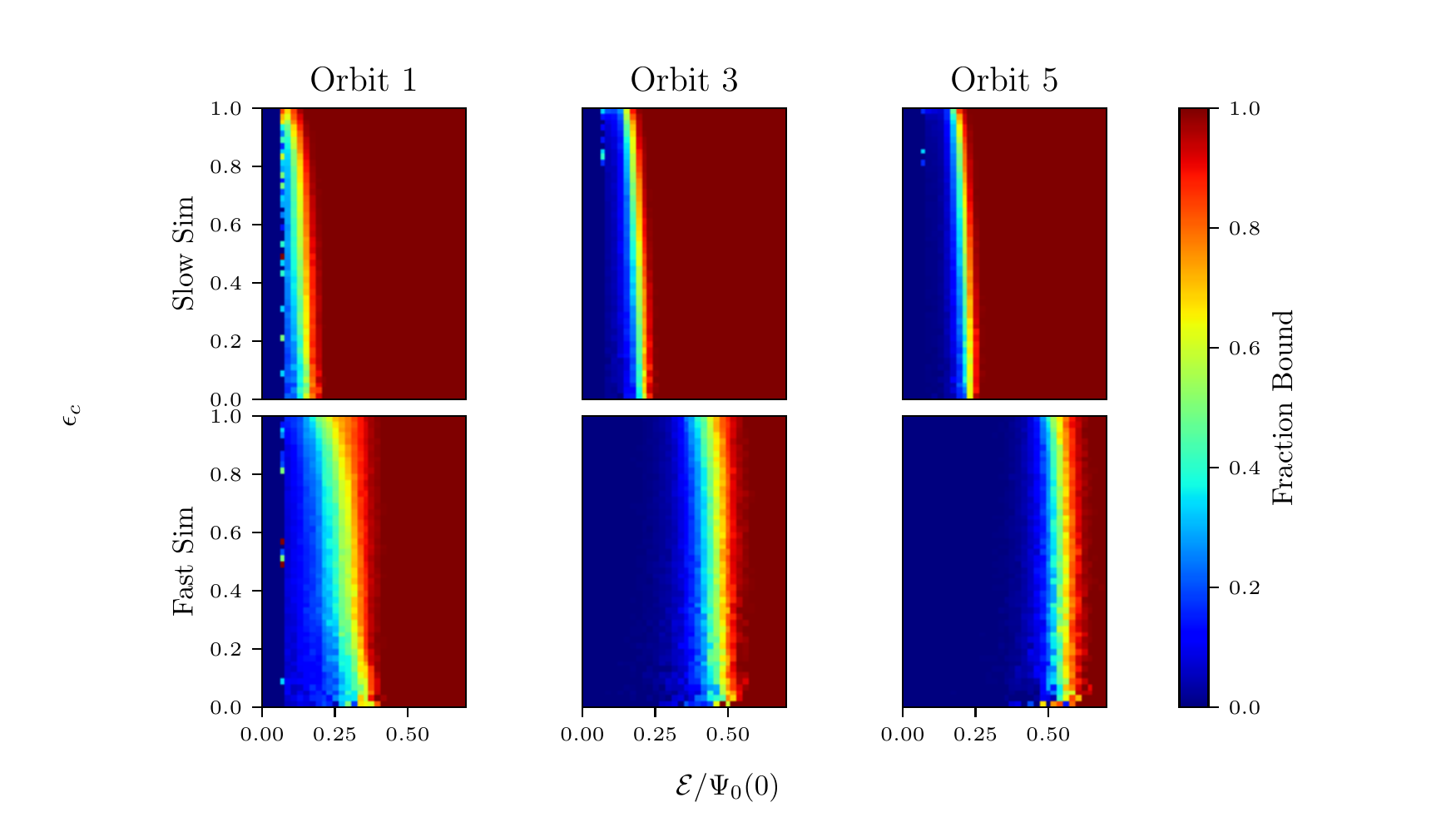}}}%
	
	\subfloat{{\includegraphics{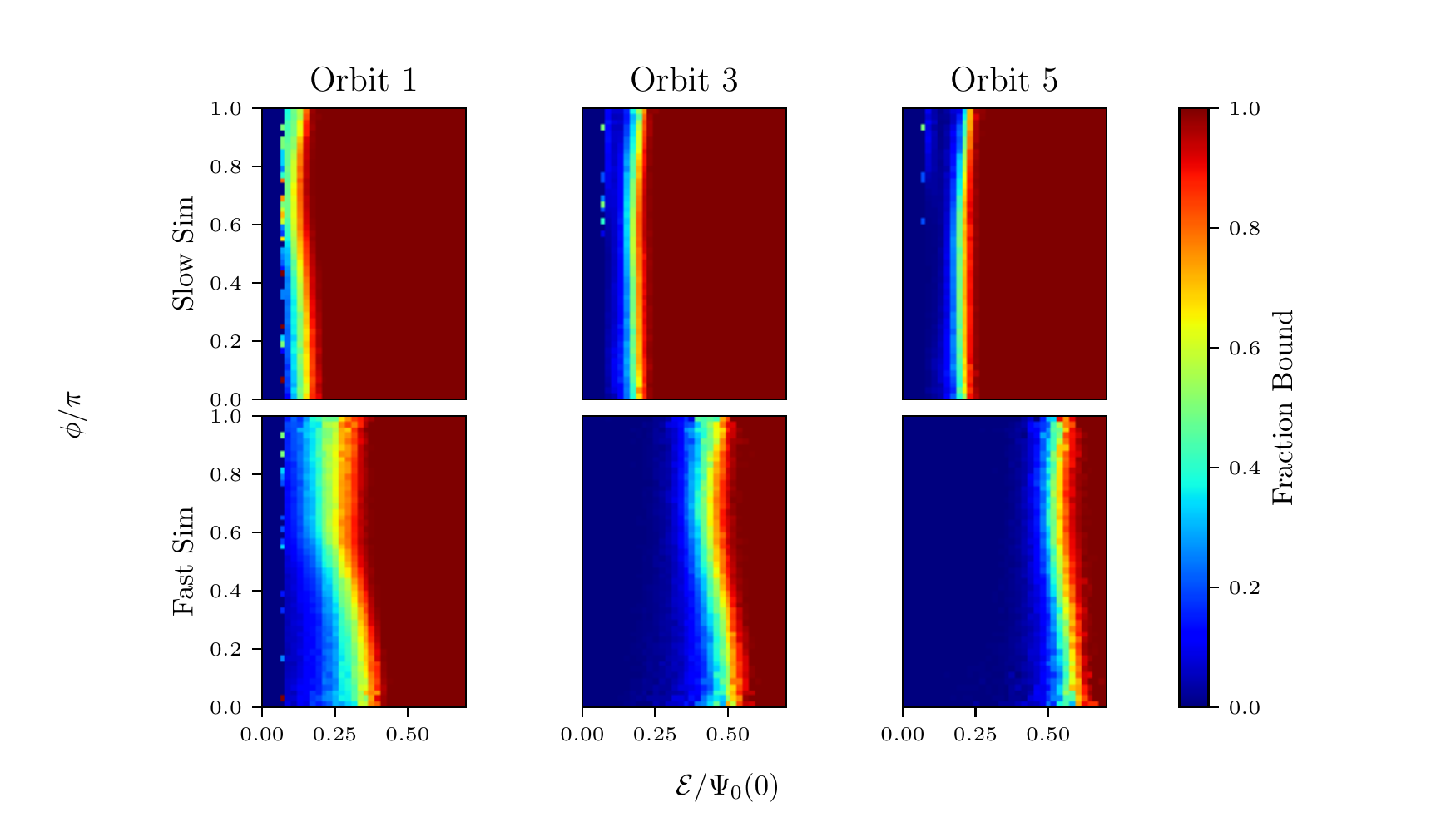}}}%
	\caption{ {\color{black} Test of whether particle removal depends on energy alone. Plots show the fraction of particles that remain bound to the remnant, as a function of the particles' initial circularity, $\epsilon_c$ (top) or inclination angle, $\phi$, (bottom) and relative energy, $\mathcal{E}$. The columns correspond to the number of orbits the subhalo has undergone, as labeled, measured at apocentre.}}
	\label{fig:EPhasePlots}
\end{figure*}

{
\color{black}
Overall, while we do find there is a dependence on particle inclination angle, overall, truncation does seem to primarily depend on particle energy, indicating assumption (ii) is valid. However, the truncation does not appear to be sharp (i.e. there is a range of truncation energy in which only a fraction of particles are removed). This might indicate `mixing'; by this we mean that individual particle energies are not conserved. Since our model removes particles once per orbit based on their instantaneous energy, while in reality they are lost over the course of the orbit, this will result in a less precise boundary in energy space. We will explore this further in Section~\ref{sec:mixing}.
}

\subsection{Scatter and mixing in energy space} \label{sec:mixing}

{
\color{black}
The final assumption of our model is that there is no variation in particle energy (or at least no change in the ordering in energy within the distribution). Since particles are not removed instantaneously, variations in energy would result in a less abrupt truncation. In this section we examine the amount of mixing in particle energies.

Fig.~\ref{fig:EPhasePlots} suggests that there may be a less abrupt truncation in the Fast Simulation compared to the Slow Simulation}. In the previous section, we explored whether this could be described by a secondary particle dependence on angular momentum or particle inclination, but neither seemed to account for the difference. Therefore, we expect it can be described by mixing; to examine this further, we plot histograms of the original energies and angular momenta removed at each time step in Fig.~\ref{fig:Mfrac_vs_E}. The slopes of the histograms become shallower at each successive orbit, but the fractional change $\Delta \mathcal{E}/\mathcal{E}_T$ is roughly constant (roughly 0.1 for the Slow Simulation, and 0.3 for the Fast Simulation), where $\Delta \mathcal{E}$ is the approximate width of the slope.

\begin{figure}
	\includegraphics[clip=true,width = \columnwidth]{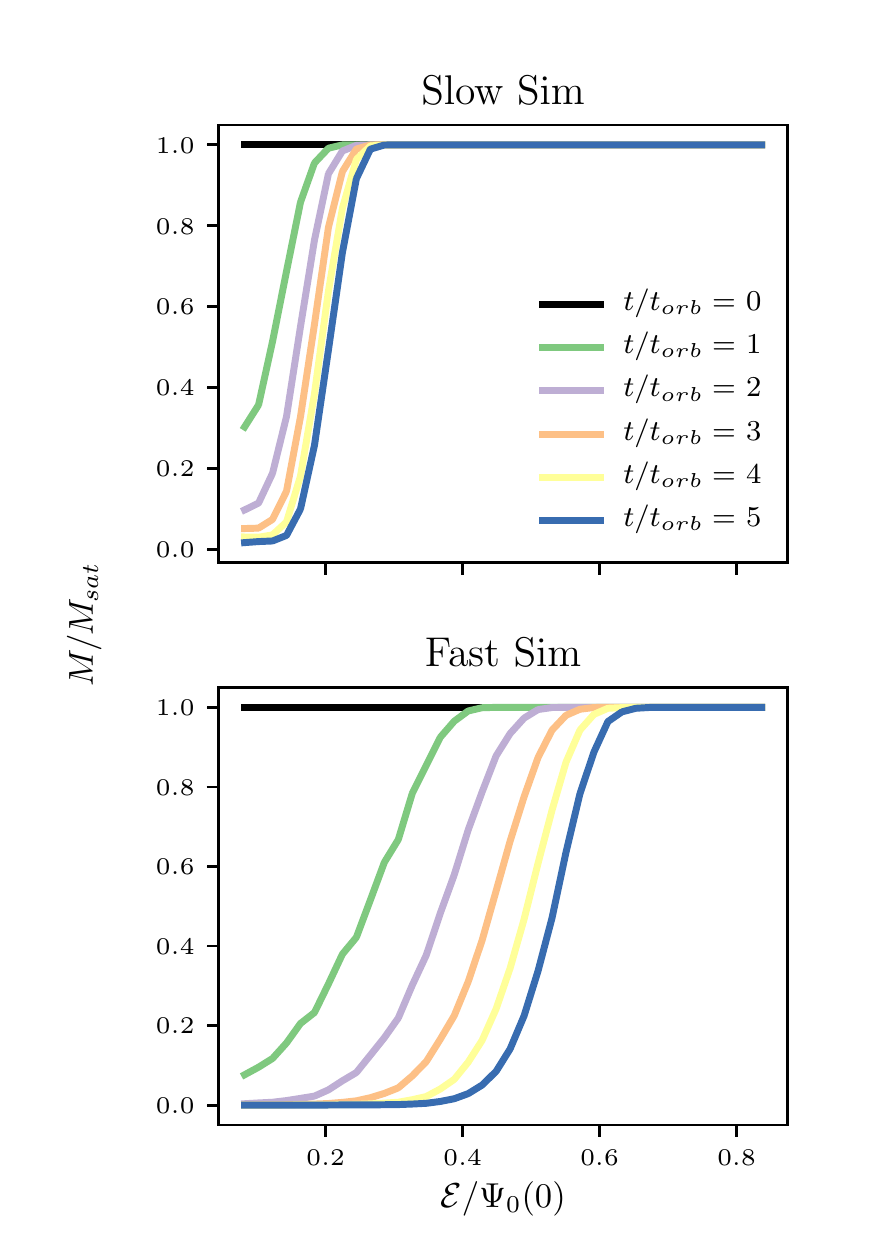}.
	\caption{The fraction of particles that remain bound to the remnant  {\color{black} at each apocentric passage as a function of their initial relative binding energy, $\mathcal{E}$. This figure demonstrates that the truncation is not as sharp in the Fast Sim case; our model implicitly assumes that these should be Heaviside step functions.}}
	\label{fig:Mfrac_vs_E}
\end{figure}

It should be noted that so far we have labeled particles by their \emph{initial} energy and angular momentum, assuming that changes in these quantities (beyond a global shift in energy) are negligible. In practice, however, these properties will change with time. In Fig.~\ref{fig:Mixing} we show the change in individual particle energies, angular momenta, radii and velocities. The changes in angular momentum and radius are roughly symmetric about zero. The binding energy, $\mathcal{E}$ decreases with each orbit; this is because the mass, and thus the total potential energy, of the system is decreasing. For the velocity distributions, there appear to be two populations in the fast simulation (i.e. there is an extra bump around $\Delta v = 0.6$) this second population is likely particles that are temporarily still bound but are leaving system. Overall, for all four quantities, the Fast Simulation has wider distributions, indicating that there is more change in particle parameters in this case.

\begin{figure*}
	\includegraphics[]{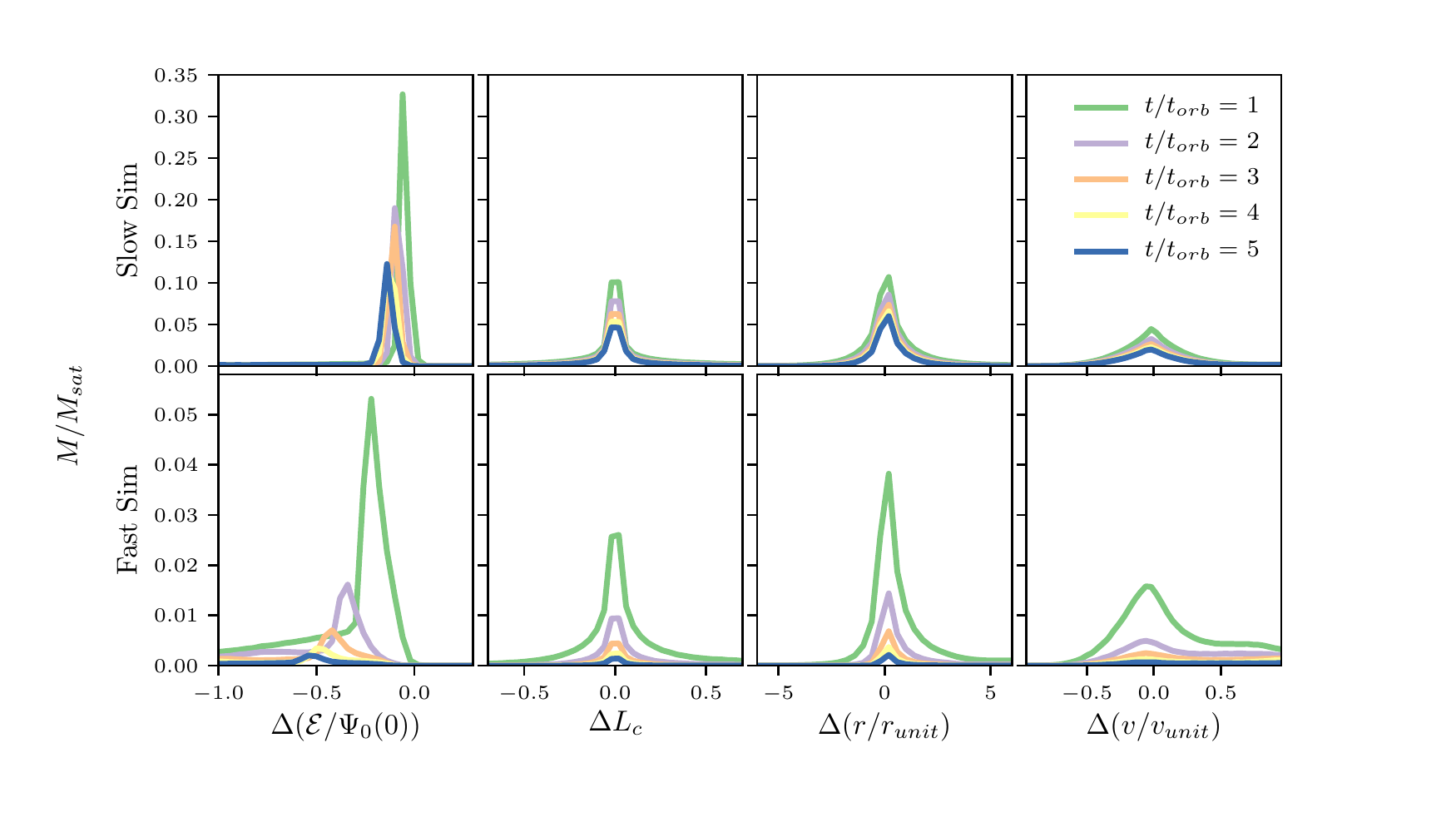}.
	\caption{{\color{black} The distribution in changes (or `mixing') in energy, angular momentum, radius and velocity of individual particles for the Slow Simulation (top) and Fast Simulation (bottom).}}
	\label{fig:Mixing}
\end{figure*}

Finally, Fig.~\ref{fig:EPhaseTri} shows the fraction of particles removed in $\epsilon_c$--$\mathcal{E}$ space, but instead of only considering the initial energy/circularity of each particle, we recalculate these values at each orbit; this way the time evolution of the location of removed particles in phase space can be examined. For both the Slow (top) and Fast (bottom) Simulations, particles are removed mainly based on energy, as shown previously. Each successive row then shows where these particles were in $\epsilon_c$--$\mathcal{E}$ space at one orbital period earlier. Interestingly, there is still a fairly discrete slice in each case; considered at successively earlier orbits, the particles that will be removed have mainly just shifted $\mathcal{E}$ (this is due to the system losing mass and thus potential energy), though there is some scatter, particularly for the Fast Simulation case. Overall, Fig.~\ref{fig:EPhaseTri} confirms that tidal stripping can mainly be described by removing particles based on their initial energies.

\begin{figure*}
	\centering
	\subfloat{{\includegraphics{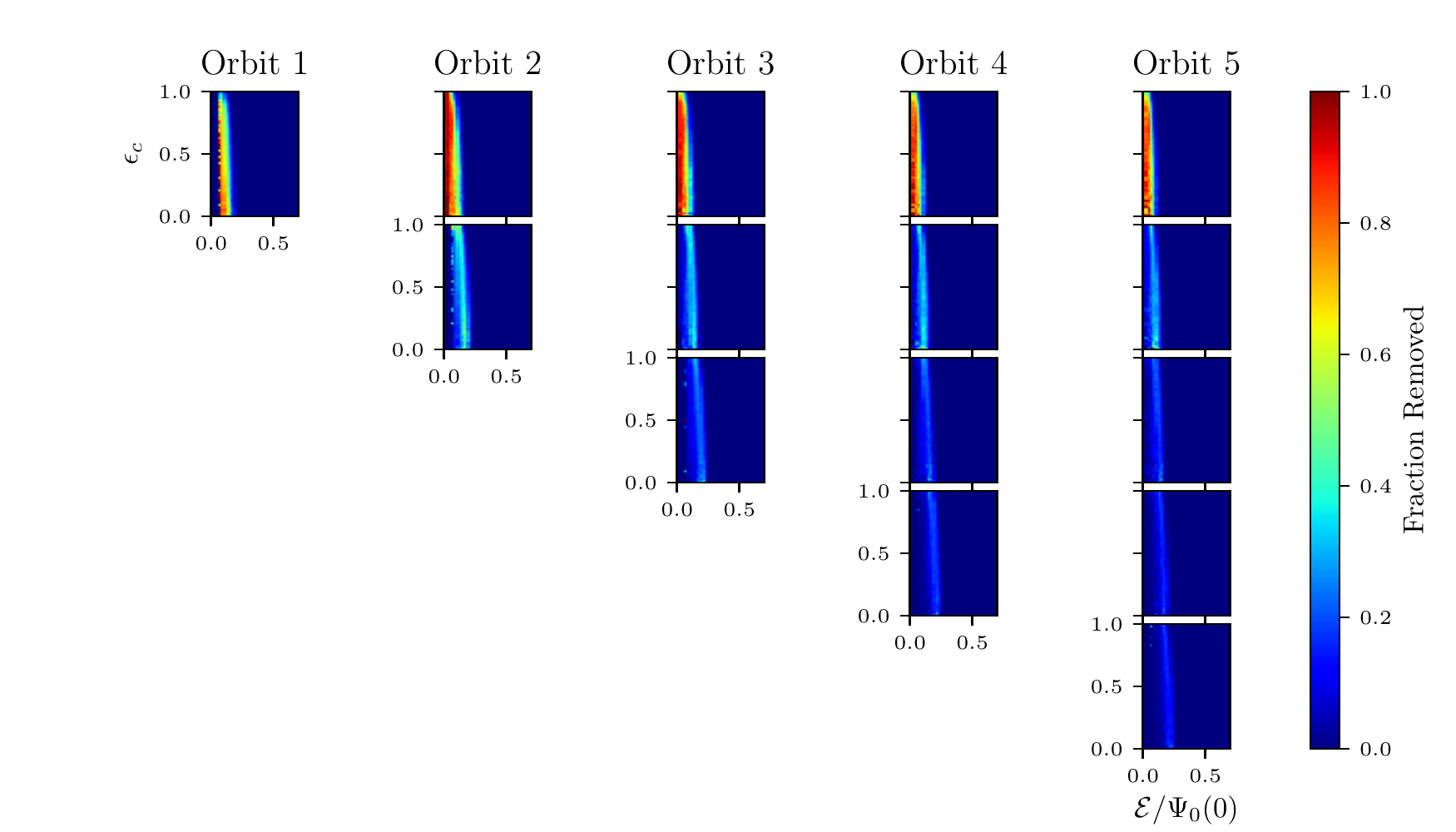}}}%
	
	\subfloat{{\includegraphics{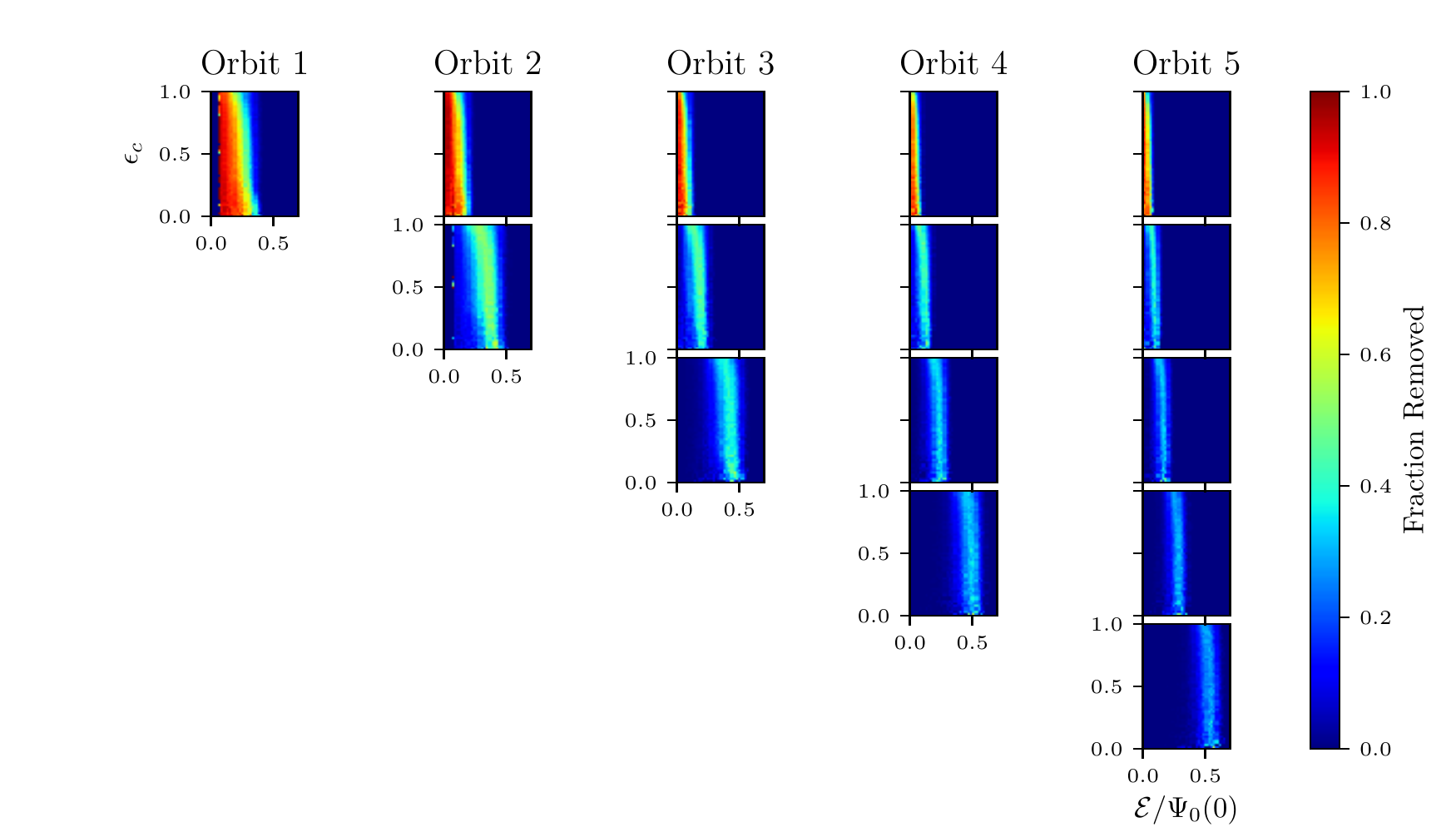}}}%
	\caption{The fraction of particles removed in $\epsilon_c$--$\mathcal{E}$ space for the Slow Simulation (top) and Fast Simulation (bottom). {\color{black}Unlike Fig.~\ref{fig:EPhasePlots}, here we consider how the energy of the particles evolve in time, which should account for some of the mixing of particle energies.} In each $n=1$-$5$ row, the particles are labeled by their angular momentum and energy $n$ orbits earlier.}
	\label{fig:EPhaseTri}
\end{figure*}

\subsection{Conclusions}

While comparing the Fast and Slow Simulations, we found that the Fast Simulation remnants are more prolate at pericentre, and more anisotropic than the Slow Simulation remnants. Both remnants do seem to be roughly in equilibrium at apocentre, except for a small number of particles at large radii. { \color{black} The largest difference between the two simulations is that the energy truncation is less sharp, as shown in Fig.~\ref{fig:Mfrac_vs_E}). Since most particles are removed during a brief interval around pericentric passage (particularly for the Fast Simulations), 
variations in energy at this time will result in an apparent blurring of the energy truncation as evaluated at the next apocentre. } We conclude that this `mixing' in energy space is likely the main cause of discrepancies between the energy-truncation model and simulations of rapid mass-loss systems.

\section{A full description of tidal mass loss} \label{sec:massloss}

{\color{black} We have shown that while the assumptions of our energy-truncation model from Paper I do not fully capture the complexity of tidal stripping, they represent a reasonable first approximation. We will now build on this approach to develop a full model of tidal mass loss. Specifically, in this section we will add an estimate of the mass loss rate, and show that matches our simulation results without any need for free parameters.} 

Conceptually, the model is very straightforward; at each pericentre mass is removed outside of a defined tidal radius, and the system then relaxes to a new equilibrium state by apocentre, achieving a new density profile that can be predicted from an energy truncation. The tidal radius of this new profile is then recalculated at the next pericentric passage, and the process is repeated.

\subsection{Mass loss model} \label{sec:masslossmodel}

Following the Jacobi model for tidal mass loss on a circular obit  \citep{binney,taylor2001a}, we can define an approximate tidal radius $r_{\rm lim}$ for a stripped system in terms of relative densities: 
\begin{equation}
\bar{\rho}_{\rm sat} (r_{\lim})= \eta \bar{\rho}_{H}(r_p) \,\,\,,
\end{equation}
where $\bar{\rho}_{\rm sat} (r_{\rm lim})$ is the mean density of the satellite interior to some radius $r_{\rm lim}$ (this is generally referred to as the tidal radius, but we will avoid this terminology so not to confuse it with the truncation radius, $r_t$, defined in Section~\ref{sec:model}), and $\bar{\rho}_{H}(r_p)$ is the interior density of the host halo within the pericentric radius. The constant, $\eta$, depends on the definition of $r_{\rm lim}$, and can be calculated analytically.

As summarized in \cite{vandenbosch2018}, depending on model assumptions, many different versions of $r_{\rm lim}$ are used in the literature. {\color{black} Each will result in a  different definition} of $\eta$; e.g., the well known Roche and Jacobii limits have $\eta$ values of 2 and 3, respectively. In general, theoretical calculations of $r_{\rm lim}$ all make a number of simplifying assumptions, such as circular orbits or the distant-tide approximation (that the distance between the satellite and host is much larger than the size of the subhalo). 

In this work we will consider two commonly used values for $\eta$, which we term $\eta_1$ and $\eta_2$; the first \citep{tormen1998}, considers the satellite and host as extended bodies, but neglects the centrifugal force, while the latter \citep{king1962} includes it:
\begin{equation}
\begin{aligned}
\eta_1 &= 2 - \dfrac{{\rm d} \ln M}{{\rm d} \ln r} \\
\eta_2 &= \dfrac{\omega^2}{\omega_c^2} - \dfrac{1}{\omega_c^2} \dfrac{{\rm d}^2 \phi}{{\rm d}r^2} = \dfrac{r_a^2v_a^2}{GM r} +2-  \dfrac{4 \pi r^3\rho}{M} \,\,\, ,
\end{aligned}
\end{equation}
where $M$ and $r$ are the {\color{black}enclosed} mass and radius of the host halo, $\omega= |\mathbf{v} \times \mathbf{r}|/r^2$ is instantaneous angular velocity of the satellite, $\phi$ is potential of host and $\omega_c^2 \equiv GM/r^3$ is the angular velocity of a circular orbit. {\color{black} For a spherically symmetric system, the potential is related to the enclosed mass by ${\rm d}\phi/{\rm d} r =- GM/r^2$ }. While these definitions are only valid for circular orbits they are commonly applied to eccentric orbits, by calculating an instantaneous value and/or a value at pericentre \citep[e.g.][]{king1962, taylor2001a, penarrubia2005, vandenbosch2018}.
Though both $\eta_1$ and $\eta_2$ assume circular orbits, the former actually works better in case of radial orbits since the extra centrifugal term is zero.  

For our mass loss model, we define an effective $\eta$, $\eta_{\rm eff}$,  by averaging the instantaneous value of $\eta_2$ over the entire orbit:
\begin{equation} \label{eq:etaeff}
\eta_{\rm eff} = \dfrac{1}{t_{orb}} \int_0^{t_{orb}} \left(\dfrac{r_a^2v_a^2}{GM r} +2-  \dfrac{4 \pi r^3\rho}{M}  \right) dt \,\,\, .
\end{equation}
As we show later, this definition gives mass loss predictions that are bracketed by those from $\eta_1$ and $\eta_2$ definitions.

Given this definition of the tidal limit, the full model for mass loss is as follows: 
\begin{enumerate}
	\item calculate the mean density of the host interior to the pericentre of the orbit, $\bar{\rho}_H (r_p)$
	\item {\color{black} calculate the enclosed density of the satellite,  according to $\bar{\rho}_{\rm sat} = \eta \bar{\rho}_H (r_p)$}
	\item calculate the bound mass; i.e. the mass within radius $r_{\rm lim}$  {\color{black}given $\bar{\rho}_{\rm sat}  = \bar{\rho}_{\rm sat} (m(<r_{\rm lim})) $ (note that the radius, $r_{\rm lim}$ does not need to be explicitly calculated)}
	\item update the profile by lowering the distribution function until the remaining bound mass matches the value calculated in (ii)
	\item repeat these last two calculations for each pericentric passage.
\end{enumerate}

\subsection{Results}

Fig.~\ref{fig:MassLossPredictionEta} shows how our full mass loss model compares to the simulation results. The simulations are sorted by increasing circularity $\epsilon_c$. As expected, $\eta_1$ tends to work better for less circular orbits, while $\eta_2$ works better for more circular orbits. In general, $\eta_1$ under-predicts the rate of mass loss, while $\eta_2$ over-predicts it. Our proposed value of $\eta$, $\eta_{\rm eff}$, works very well in most cases. All three definitions severely under-predict mass loss in Simulations 9 and 10; however these two simulations have the smallest mass ratios (i.e.~the largest satellites relative to the scale of the host), so the distant-tide approximation in calculating $\eta$ is less valid. 

\begin{figure*}
	\includegraphics[]{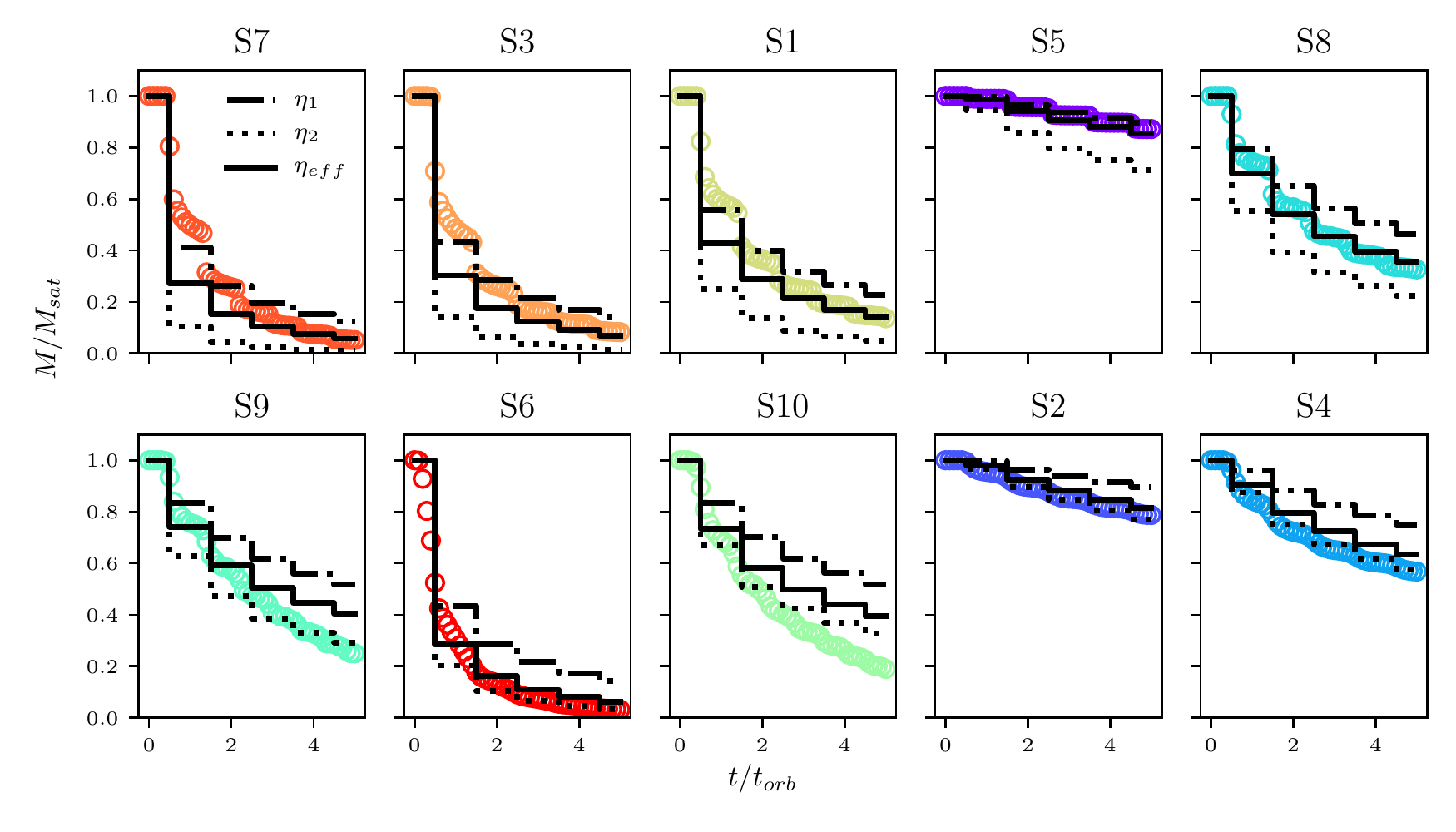}.
	\caption{Predicted bound mass versus time. The simulations are shown with points, and our model with lines, for three different definitions of $\eta$, {\color{black} as described in Section~\ref{sec:masslossmodel}}. Simulations are sorted from left to right, top to bottom in terms of increasing circularity.}
	\label{fig:MassLossPredictionEta}
\end{figure*}

Additionally, we show mass loss predictions using different profile models in Fig.~\ref{fig:MassLossPrediction_hayashi}, using $\eta=\eta_{\rm eff}$. {\color{black} These were calculated as described in Section~\ref{sec:masslossmodel}, but replacing step (iii) with the appropriate density profile model from Appendix~\ref{sec:models}}. All models agree fairly well for the first few orbits, but then begin to deviate. Overall, our model predicts the least amount of mass loss; often the other models start to over-predict mass loss after a few orbits. For some of the orbits (e.g. S4, S9, S10), our model may under-predict mass loss slightly, but generally the agreement is good.

\begin{figure*}
	\includegraphics[]{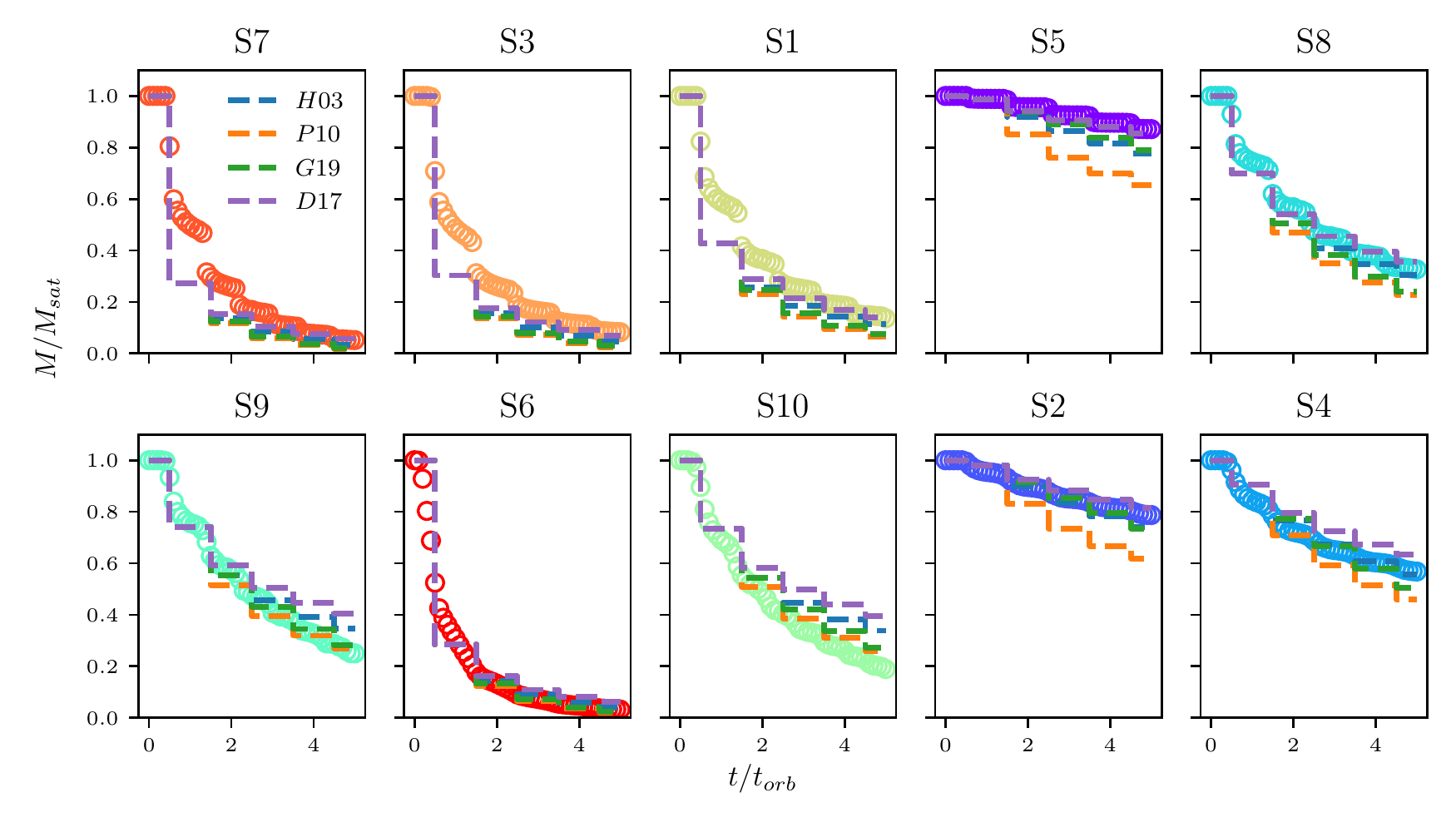}.
	\caption{Bound mass predictions versus time, using $\eta = \eta_{\rm eff}$ (Equation~\eqref{eq:etaeff}). Simulations are sorted from left to right, top to bottom in terms of increasing circularity. The updated profiles in the mass calculation  (step (iii) in Section~\ref{sec:masslossmodel}) were from either Paper I (D17), \protect\citep{hayashi2003} (H03), \protect\citep{penarrubia2010} (P10) or \protect\citep{green2019} (G19).}
	\label{fig:MassLossPrediction_hayashi}
\end{figure*}

Finally, we check these predictions still fit the profile well in Fig.~\ref{fig:NFW_compare_new} and find that the accuracy is similar in both cases. For the Slow Simulation, the fits are about the same, while for the Fast Simulation the theoretical prediction for $\mathcal{E}_T$ prefers a lower satellite mass; this is not surprising, since as discussed in Section~\ref{sec:lim}, the Fast Simulation contains a population of particles that are temporarily bound, but leaving the system. In either case, this prediction does not improve the residuals compared to the direct fit to the total mass. This demonstrates there is some genuine difference between the Fast Simulation and the energy truncation model, and the disagreement between them is not merely due to the method used to fit the simulation.

\begin{figure*}
\includegraphics[]{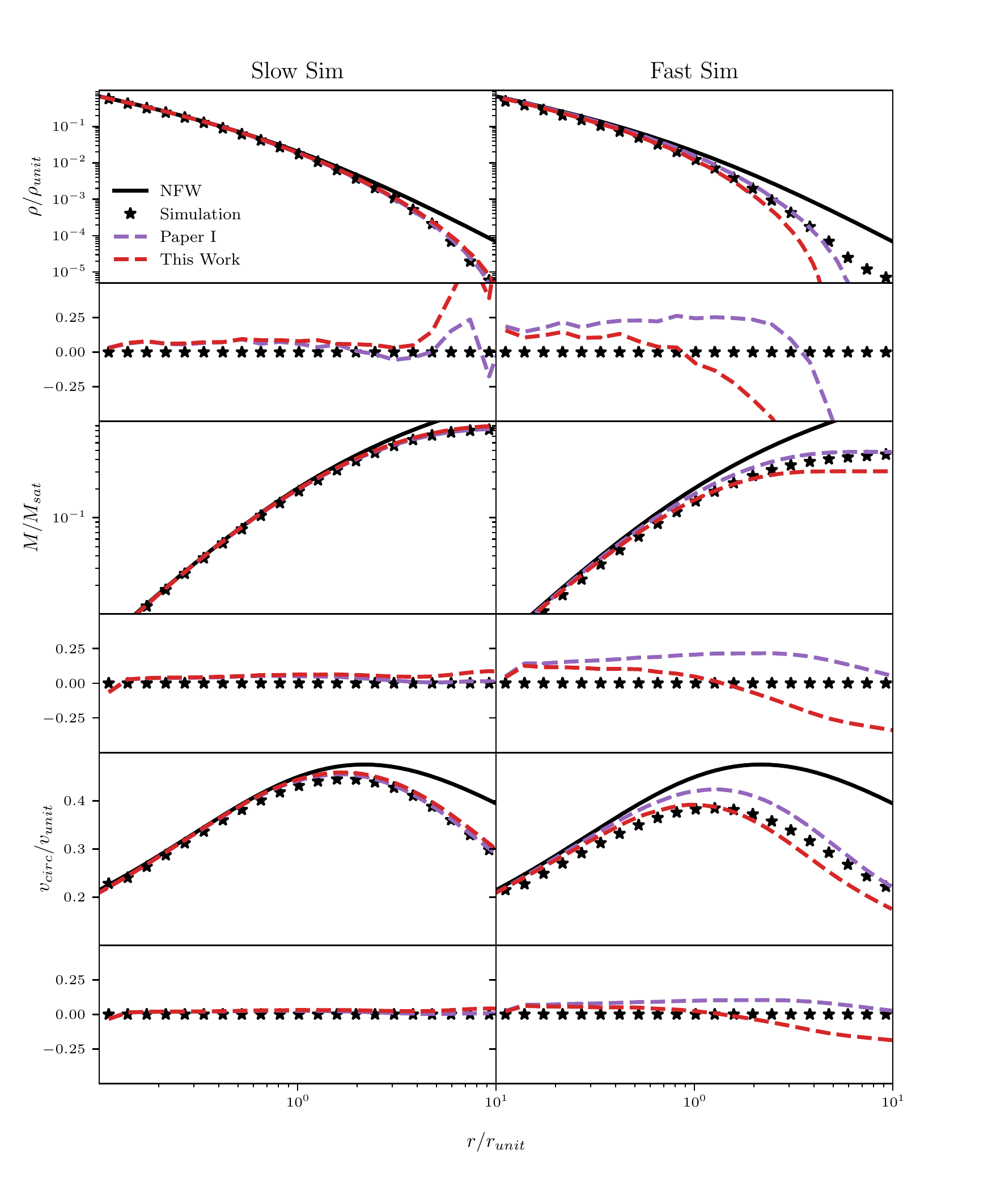}
\caption{Comparison of density, mass and velocity profiles of the bound satellite remnant after one orbit for different methods of fitting $\mathcal{E}_T$ to the Slow Simulation (left) and Fast Simulation (right). Simulation results are shown with points, and the model is shown with lines. The thick dashed curve shows the original, un-truncated NFW profile. The residuals in the fits are calculated as ($y_{\rm model}/y_{\rm simulation}-1$), where $y$ is the density, mass or circular velocity profile. In Paper I, the model was fit using the total bound mass fraction {\color{black} calculated from the simulations}, while in this work we predict the bound mass fraction as outlined in Section~\protect\ref{sec:masslossmodel}.}
\label{fig:NFW_compare_new}
\end{figure*}

We have only considered a simple orbit-averaged model, in which all mass loss occurs instantaneously at pericentre, but clearly this is a simplification. Alternatively, it is common practice to make a continuous model by dividing the orbital period into discrete steps and assuming a fraction of the mass outside the tidal radius is lost at each of these steps, according to a characteristic time scale for mass loss \citep{taylor2001a}. Following \cite{vandenbosch2018}, the mass-loss rate can be expressed as
\begin{equation}
\dot{m} = \frac{\alpha}{ \tau_{\rm orb}} m(<r_{\rm lim}) \, ,
\end{equation}
where $r_{\rm lim}$ is instantaneous tidal radius and $ \tau_{\rm orb} = 2 \pi/ \omega$ is the instantaneous angular velocity. However, there is some debate over the value of $\alpha$. While $\alpha$ was originally given a value of $1$ \citep[e.g.][]{taylor2001a, taffoni2003, zentner2003, taylor2004}, calibrations to cosmological simulations suggest a range of $\alpha$ values from $2.5$-$6$ \citep{zentner2005,diemand2007,pullen2014}. Testing this approach, we find that our results are consistent with $\alpha \approx 2.5$. {\color{black} However, since we do not have a physical argument for setting $\alpha=2.5$, we do not currently include continuous mass loss in our model}. Furthermore, our profile model assumes the system is in equilibrium, so updating the profile during fractions of the orbital period is likely problematic.

\section{Discussion} \label{sec:conc}

In Paper I, we proposed an energy-truncation model to describe {\color{black} how tidally stripped haloes lose mass}, and showed that it reproduced the structure and distribution functions seen in isolated simulations of NFW subhalo evolution. In this work, we have shown that this model predicts the evolution of the density profile with an accuracy similar to empirical models \citep[e.g.][]{hayashi2003, penarrubia2010, green2019}, but has the advantage of being physically motivated, allowing for a theoretical prediction of density evolution at very small radii. Examining the model assumptions in more detail, we have determined that {\color{black} as well as an overall shift, there is some scatter in individual particle energies from one orbit to the next}, particularly in cases of rapid mass loss. This results in less abrupt energy truncation than is assumed in our model, and is probably the main factor limiting its accuracy.} 

{\color{black} The energy truncation approach introduced in Paper I did not predict the truncation energy $\mathcal{E}_{\rm T}$, and thus the mass-loss rate was not specified.} In this paper, we estimate the mass-loss rate by assuming that the profile is truncated to some radius $r_{\rm lim}$ (which can be defined in terms of the mean enclosed density). The system is then assumed to relax and rearrange itself into a new, lowered DF. We expressed $r_{\rm lim}$ in terms of a mass loss parameter $\eta$, and found that common definitions of $\eta_1$ and $\eta_2$ worked well at predicting the bound mass fraction in the case of radial and circular orbits, respectively. We defined a new $\eta$, $\eta_{\rm eff}$, which takes the average $\eta_2$ value over the whole orbit, and found it is a good predictor of the bound mass fraction, as long as the distant-tide approximation is valid. Combining energy truncation with this estimate of the mass-loss rate, we have obtained a complete model {\color{black} for tidal mass loss that has no free parameters.}

One advantage to more physically motivated models such as the one we propose is that they may help distinguish between real processes and numerical artifacts. For example, there has been some debate in the literature over whether subhaloes ever fully disrupt; a recent paper by \cite{vandenbosch2018} offers a convincing argument that most subhalo disruption occurs when the tidal radius is smaller than the softening length, and is thus numerical in nature. Further, \cite{errani2019} recently performed isolated simulations in which they repeatedly reconstruct the density cusp of the merging system and conclude  that cuspy dark matter haloes can never be completely disrupted in smooth tidal fields. This is consistent with our model; we predict that at each passage haloes will be stripped down to some interior mean density. The profile will re-arrange in such a way that the DF is the same for the most bound particles. For NFW profiles the inner density and DF diverge, and thus there will always exist a tidal radius greater than zero {\color{black} that encloses a given (finite) mean density, $\bar{\rho}_{\rm sat}$}, even at arbitrarily small bound mass fractions. 

One complication with the previous picture is that the structure of haloes at small radii is undetermined from cosmological simulations. NFW profiles are merely fits to haloes within the resolved range of simulations, so the central behaviour is unclear; isolated simulations typically assume an NFW profile for the satellite, and thus the predictions made about disruption may not be valid. There has been some work examining the effect of the central profile on subhalo evolution; for example, \cite{penarrubia2010} have studied the differences between cusped and cored profiles, and find that cored halo profiles rapidly decrease in central density, while cuspy profiles are preserved. We conclude that to make actual predictions of whether a subhalo will fully disrupt, we  need a prediction of the initial subhalo density profile {\color{black} from first principles}. In the meantime, we will explore our model predictions for alternate profile models in Paper III.
 
Not only is the central density of haloes important to halo disruption predictions, it also has implications for the dark matter annihilation boost factor \citep[e.g.][]{okoli2018, drakos2019b}, as we will explore in Paper III. We also note that our model may not capture decreased central density due to heating, though we suspect that neglecting this is valid, as  it has been found that the central regions of haloes are adiabatically shielded from heating \citep{weinberg1994a,weinberg1994b}. {\color{black} However, as shown in \cite{delos2019c}, including the effects of stellar encounters on the subhalo can cause heating of all particles in the subhalo, resulting in a change in the DF.  Incorporating the effects of these more complicated merging scenarios is a potential extension to our model.}

While the full model described in the final section of this paper predicts mass loss and structural changes without any further need for calibration, it does include a number of simplifications that may limit its accuracy. Further work is required to determine the effects of heating and shape on subhalo evolution, as well as to understand the details of mass loss. To first order, tidal truncation does seem to be strictly a function of energy. Beyond this, the angular momenta and inclinations of individual particle orbits do have some effect, as suggested in previous work \citep{read2006a}. Lastly, the 3D orientation, spin and  sphericity of the remnant may be important. {\color{black} Overall, however, we found that scatter in particle energies (mixing) likely dominates inaccuracies in the model. Slight modifications to the energy truncation model could capture this effect (e.g. allowing for a more gradual energy cut-off); this would require an additional parameter to the truncation model, and is something we will examine in future work.}

While we have assumed that mass loss is instantaneous, in reality, while some particles are instantaneously removed, other particles are lost on a slower timescale. Therefore, updating the density profile or DF over the course of the orbit requires a deeper understanding of these timescales. A further complication is that when the subhalo is not at apocentre, the assumption of pseudo-equilibrium used in our profile model may break down. This does indeed appear to happen; in fact, in our simulations we find that sometimes the tidal radius of the bound material \emph{increases} between pericentre and apocentre before the material settles down to an equilibrium state.

In this work we have only used one satellite model, on ten different orbits. Future work could explore our findings for a much larger range of simulations; the recently published DASH simulations \citep{ogiya2019} will be a useful tool for this end. The large number of simulations will also be useful for data-driven approaches to study the physical processes involved in tidal stripping; for instance principal component analysis (PCA) could be used to determine the best predictors of our model accuracy.

Overall, this work provides a simple description of subhalo evolution that does not require calibration to simulations. This is complementary to more qualitative descriptions of halo evolution \citep[e.g.][]{hayashi2003,penarrubia2010,green2019} which are useful in semi-analytic models, but are less physically motivated.


	\section*{Acknowledgements}
{\color{black}The authors would like to thank the anonymous referee for useful comments.} NED acknowledges support from NSERC Canada, through a postgraduate scholarship {\color{black} and NASA contract NNG16PJ25C}. JET acknowledges financial support from NSERC Canada, through a Discovery Grant. 
	
		\bibliographystyle{mnras}
		\bibliography{TS2_MassLoss}

	\appendix

{\color{black}
\section{Bound mass calculation} \label{sec:potential}

To determine the self-bound mass of the subhalo, we iteratively removed any particles with a negative energy. This requires calculating the potential of all the particles in the subhalo. We assumed the potential was spherically symmetric, which is a typical assumption in halo finders \citep[e.g.][]{gill2004,AHF}, as this greatly reduces the computation time. In this section, we examine how this assumption affects the results. 

\subsection{Potential calculations}
	
For a spherically symmetric system, if we treat the mass of each particle $i$ as being distributed over a shell of radius $r_i$, the potential at particle $i$ becomes:
 
\begin{equation} \label{eq:pot_spherical}
P_i \approx -Gm \left( \dfrac{N(<r_i)}{r_i} + \sum_{j=1,\\ r_j>r_i}^N \dfrac{1}{r_j} \right) \,\,\, .
\end{equation}
 
 Since this calculation involves distances $r_i$ from the centre of the subhalo, it is much faster than calculating the distances between every pair of particles, $r_{ij}$.	
 
 Due to computational limitations, it is not feasible to calculate the full potential directly; therefore, following \cite{bett2010}, we calculate the potential using only a subsample of the particles, as described in Equation~\eqref{eq:pot_bett} (another solution is to use a Barnes \& Hut octree algorithm, which was the approach in \cite{vandenbosch2018}):
 \begin{equation} \label{eq:pot_bett}
P_i = \left(\dfrac{N-1}{N_{\rm sel}-1}\right) \left(\dfrac{-Gm}{\epsilon}\right) \sum_{j=1,\\ j \neq i}^{N_{\rm sel}} -W(r_{ij}/\epsilon) \,\,\, .
 \end{equation}
Here, $N_{\rm sel}$ is the number of randomly selected particles used to approximate the entire distribution (each of these particles can be considered to have a mass $m N/N_{\rm sel}$).  As in \cite{drakos2019a},  we use $N_{\rm sel}= 5000$ particles. $\epsilon$ is the softening length used in the simulation, and $W$ is the smoothing kernel used for force calculations in \textsc{gadget-2}, given by:
 
\begin{equation}
 W(x) =
 \begin{cases}
 \dfrac{16}{3}x^2 - \frac{48}{5}x^4 + \frac{32}{5} x^5 -  \frac{14}{5}, & 0 \leq x \leq  \frac{1}{2}\\
 \frac{1}{15x} + \frac{32}{3}x^2 - 16 x^2 +\frac{48}{5}x^4\\ -\frac{32}{15}x^5 - \frac{16}{5},
 & \frac{1}{2} \leq x \leq  1\\
 -\frac{1}{x}, &  x \geq 1 \,\,\, .
 \end{cases}
\end{equation}

 \subsection{Results}
 
 We compared for the Fast and Slow Simulations; specifically we show mass loss curves in Fig.~\ref{fig:Pot_MassLoss}, density profiles in Fig.~\ref{fig:Pot_Dens} and shape measurements in Fig.~\ref{fig:Pot_Shape}.  Even though there is quite a bit of noise associated with the method used to calculate the full potential (this could be improved by repeating the method with another randomly selected subset of particles, and averaging the results), the two methods agree very well.

 \begin{figure}
 	\includegraphics[width = \columnwidth]{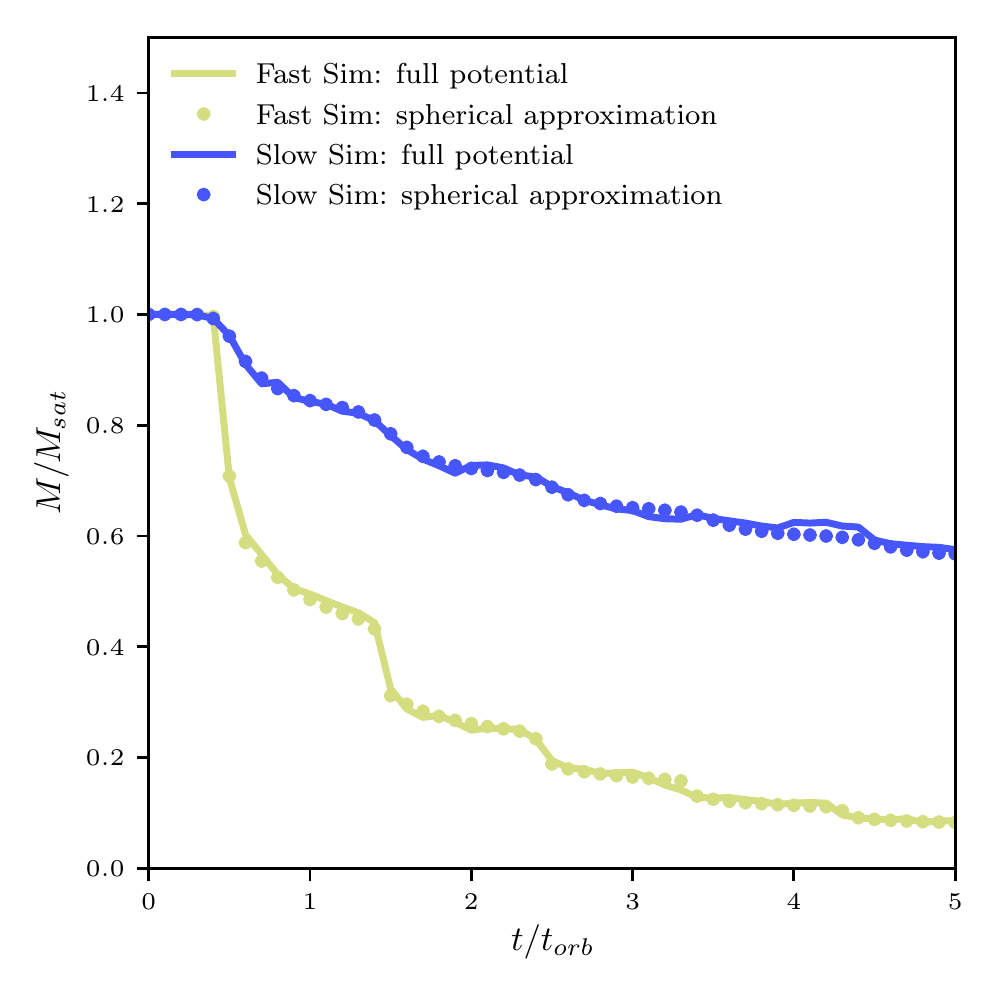}.
 	\caption{Bound mass fraction as a function of time for the Fast and Sim Simulations. The bound mass was calculated using either a spherical approximation (Equation~\protect\eqref{eq:pot_spherical}) or the full potential using a subset of particles (Equation~\protect\eqref{eq:pot_bett}).}
 	\label{fig:Pot_MassLoss}
 \end{figure}

 \begin{figure}
 	\includegraphics[clip=true,width = \columnwidth]{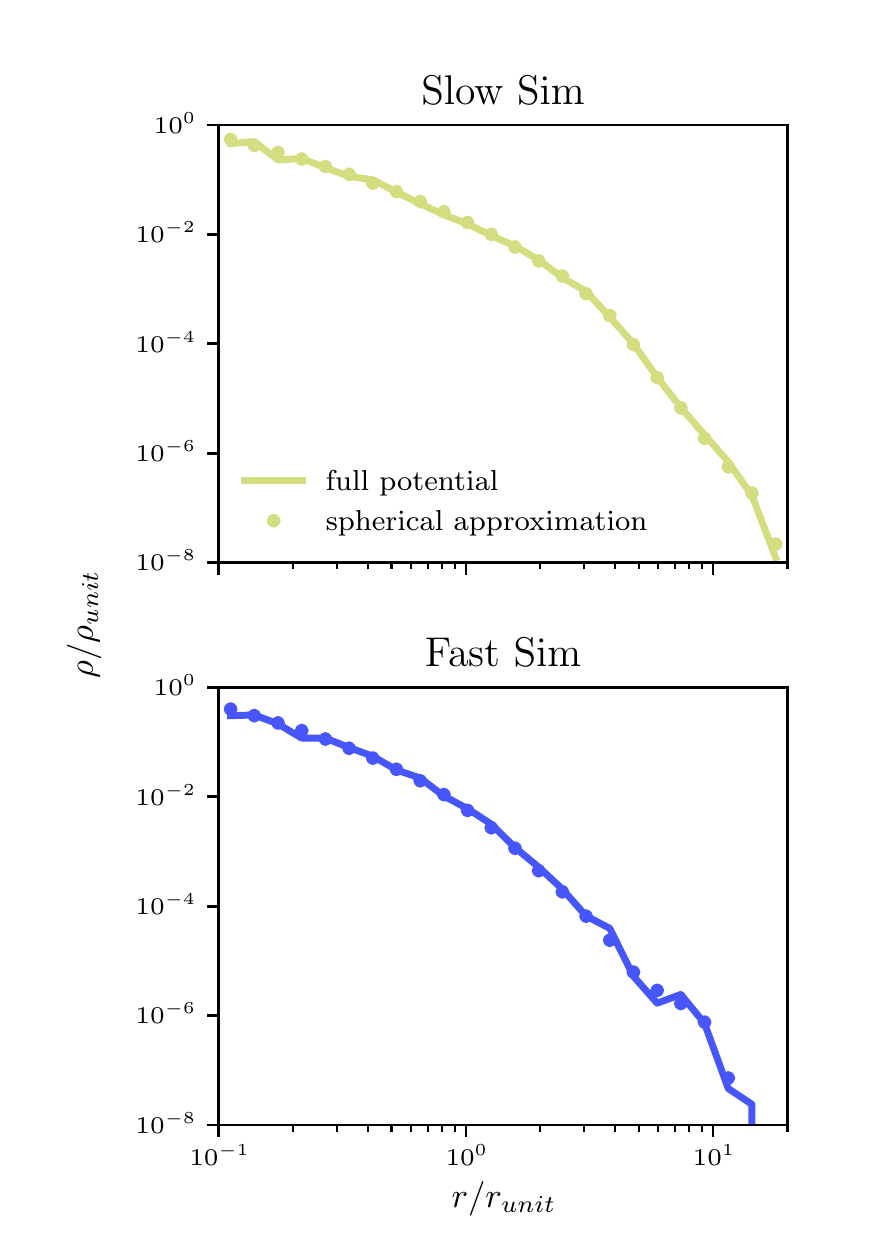}.
 	\caption{Density profiles of the Fast and Slow Simulations, measured at the third apocentric passage. The bound mass was calculated using either a spherical approximation (Equation~\protect\eqref{eq:pot_spherical}) or the full potential using a subset of particles (Equation~\protect\eqref{eq:pot_bett}).} 
 	\label{fig:Pot_Dens}
 \end{figure}

 \begin{figure}
 	\includegraphics[clip=true,width = \columnwidth]{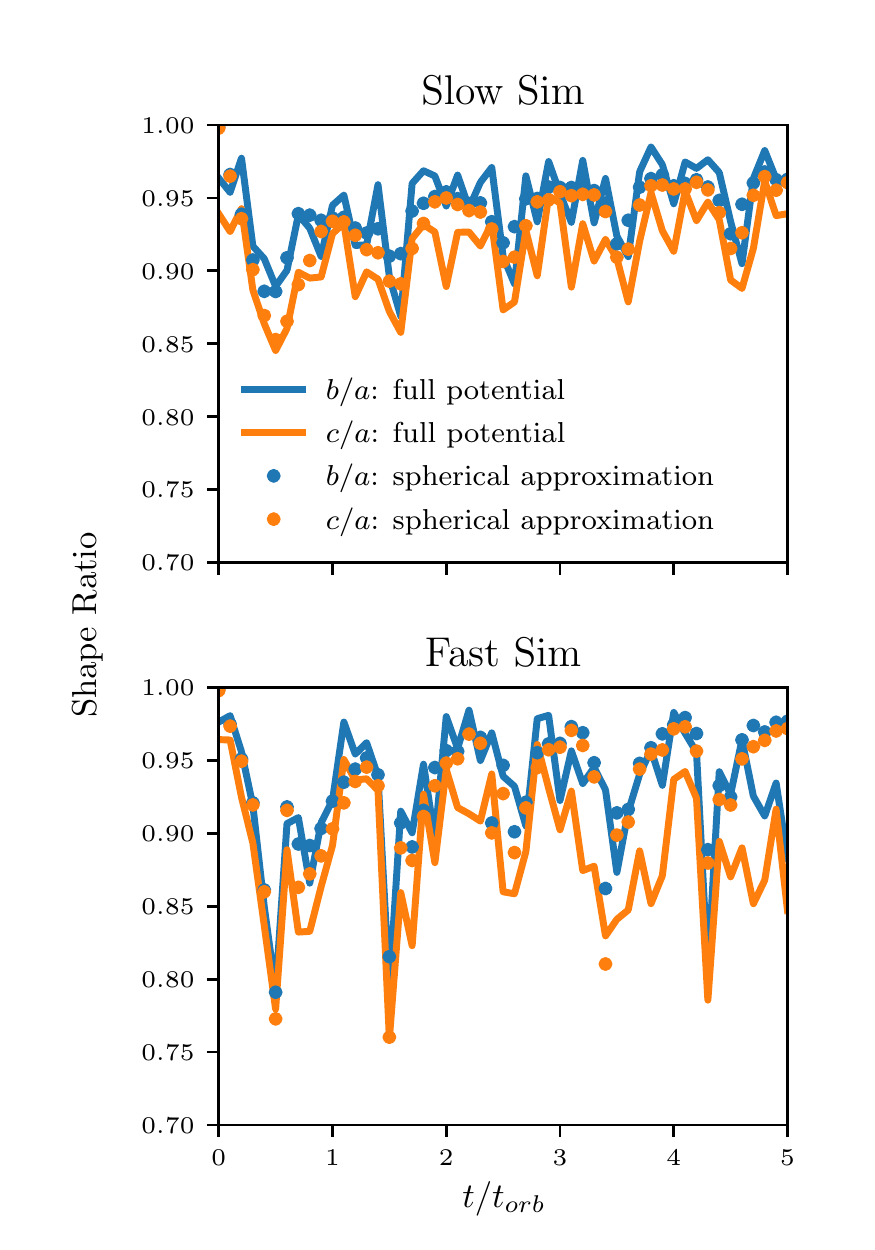}.
 	\caption{{\color{black}Subhalo shape ratios as a function of time for the Fast and Slow Simulations. The bound mass was calculated using either a spherical approximation (Equation~\protect\eqref{eq:pot_spherical}) or the full potential using a subset of particles (Equation~\protect\eqref{eq:pot_bett}).}}
 	\label{fig:Pot_Shape}
 \end{figure}

}

\section{Alternative models} \label{sec:models}
\subsection{Hayashi et al. 2003}
	
\cite{hayashi2003} proposed that tidally stripped NFW haloes have the form:
\begin{equation}  \label{eq:hayashi}
\rho(r) = \dfrac{f_t}{1 + (r/r_{te})^3} \rho_{NFW} (r) \,\,\,.
\end{equation}
The parameters $f_t$ and $r_{te}$ give a measure of the reduction in central density, and an effective tidal radius. Both of these parameters can be estimated using a single parameter---the bound mass fraction, $f_{\rm b}$, of the satellite:
	\begin{equation}
	\begin{aligned}
	\log(r_{te}/r_{\rm s}) &= 1.02 + 1.38 \log f_{\rm b} + 0.37 (\log f_{\rm b})^2\\
	\log f_t &= -0.007 + 0.35 \log f_{\rm b} \\
	& \hspace{1.5cm} +0.39 (\log f_{\rm b})^2 + 0.23 (\log f_{\rm b})^3 \,\,\,.
	\end{aligned}
	\end{equation}
Since $f_{\rm b}$ is dependent on the method used to truncate the initial NFW profile, we defined $f_{\rm b}$ to be the mass of the bound satellite compared to the mass of an untruncated NFW profile within radius $r_{\rm cut}$. 
	
\subsection{Pe\~{n}arrubia et al. 2010}

The \cite{penarrubia2010} model considers density profiles of the form:
\begin{equation}
		\rho(r) = \dfrac{\rho_0}{(r/{r_{\rm s}})^{\gamma}[1 +(r/r_{\rm s})^{\alpha}]^{(\beta - \gamma)/\alpha}} \,\,\, .
\end{equation}
For this paper, we only consider NFW profiles, which have $(\alpha,\beta,\gamma)=(1,3,1)$. The procedure for determining the evolution of the profile is outlined in the following steps.

\subsubsection{Step 1}
First, the bound mass fraction of the satellite as calculated. As before, we calculated $f_{\rm b}$ as the mass of the bound satellite compared to the mass of an untruncated NFW profile within radius $r_{\rm cut}$. If $f_{\rm b} \leq 0.9$, then $(\alpha,\beta,\gamma) \rightarrow (\alpha,5,\gamma)$.
	
\subsubsection{Step 2}
Secondly, $v_{\rm max}$ and $r_{\rm max}$  (the peak of the circular velocity curve and corresponding radius) can be calculated empirically. As originally described in \cite{penarrubia2008b}, the evolution of various subhalo structural parameters can be described by
\begin{equation}
g(f_{\rm b}) = \dfrac{2^\mu f_{\rm b}^\eta}{(1+f_{\rm b})^\mu} \,\,\, .
\end{equation}
In \cite{penarrubia2010} they showed that for an NFW profile, the best fit parameters are $(\mu,\eta)=(0.4,0.3)$ for $g(f_{\rm b})=v_{\rm max}/v_{\rm max}(0)$, and $(\mu,\eta)=(-0.3,0.4)$ for $g(f_{\rm b})=r_{\rm max}/r_{\rm max}(0)$.

\subsubsection{Step 3}

The scale radius, $r_{\rm s}$ can be calculated from the circular velocity profile, $V_c = \sqrt{GM(r)/r}$, by noting that $dV_c/dr (r_{\rm max}) = 0$. Then,
\begin{equation}
\dfrac{dM}{dr}\bigg|_{r = r_{\rm max}} = \dfrac{M(r_{\rm max})}{r_{\rm max}}
\end{equation}
	
For an NFW ($(\alpha,\beta,\gamma)=(1,3,1)$) profile:
\begin{equation}
\begin{aligned}
M(r) &= 4 \pi \rho_0 r_{\rm s}^3 \left[ \ln\left( \dfrac{r_{\rm s} +r}{r_{\rm s}}\right) -\dfrac{r}{r_{\rm s}+r} \right] \\
\dfrac{dM(r)}{dr}&= 4 \pi \rho_0 r_{\rm s}^3 \dfrac{r}{(r+r_{\rm s})^2}
\end{aligned}
\end{equation}
	
while, for a $(\alpha,\beta,\gamma)=(1,5,1)$ profile:
\begin{equation}
\begin{aligned}
M(r) &= \dfrac{2}{3} \pi \rho_0 r_{\rm s}^3  \dfrac{(r+3 r_{\rm s}) r^2}{(r+r_{\rm s})^3}  \\
\dfrac{dM(r)}{dr} &= 4 \pi \rho_0 r_{\rm s}^5 \dfrac{r }{(r+r_{\rm s})^4}
\end{aligned}
\end{equation}
	
Overall, the scale radius of a stripped NFW profile is given by: 
\begin{equation}
\begin{aligned}
\ln\left( \dfrac{r_{\rm s} +r_{\rm max}}{r_{\rm s}}\right) &= \dfrac{r_{\rm max}^2}{(r_{\rm max}+r_{\rm s})^2} + \dfrac{r_{\rm max}}{r_{\rm s}+r_{\rm max}} \,\,\, \\& 
\hspace{3cm} \mbox{for} \,\,\, f_{\rm b} > 0.9\\
r_{\rm s} & =\left( \dfrac{2}{3} + \dfrac{\sqrt{7}}{3} \right) r_{\rm max}  \,\,\, \mbox{for} \,\,\, f_{\rm b} \leq 0.9
\end{aligned}
\end{equation}
	
\subsubsection{Step 4}
	
Finally, $\rho_0$ can be calculated by using $v_{\rm max} = v_{\rm circ}(r_{\rm max})$:
\begin{equation}
\begin{aligned}
\rho_0&=\dfrac{v_{\rm max}^2r_{\rm max}}{G} \left(4 \pi  r_{\rm s}^3 \left[ \ln\left( \dfrac{r_{\rm s} +r_{\rm max}}{r_{\rm s}}\right) -\dfrac{r_{\rm max}}{r_{\rm s}+r_{\rm max}} \right]  \right)^{-1}\,\,\,, \\& \hspace{5cm} \mbox{for} \,\,\, f_{\rm b} > 0.9\\
\rho_0&=\dfrac{v_{\rm max}^2r_{\rm max}}{G} \left(\dfrac{2}{3} \pi r_{\rm s}^3  \dfrac{(r_{\rm max}+3 r_{\rm s}) r_{\rm max}^2}{(r_{\rm max}+r_{\rm s})^3} \right)^{-1}\,\,\,, \\& \hspace{5cm} \mbox{for} \,\,\, f_{\rm b} \leq 0.9 \,\,\,.
\end{aligned}
\end{equation}

\subsection{Green and van den Bosch, 2019}

In their recent paper, \cite{green2019} have provided 
an update to the empirical model proposed by \cite{hayashi2003}. While they still only consider NFW profiles, they vastly improve the accuracy of the model by using the publically available DASH simulations \citep{ogiya2019}. This library of simulations are carefully initialized by sampling the DF (rather than the Maxwellian approximation used in \cite{hayashi2003}). Additionally they consider a wide range of parameter space, including different satellite concentrations, $c$.

The model is given by:
\begin{equation}  
\rho(r) = \dfrac{f_t}{1 + \left( r\left[ \dfrac{r_{\rm vir} - r_{te}}{r_{\rm vir} r_{te}} \right]\right)^{\delta}} \,\,\,\rho_{NFW} (r) \,\,\,,
\end{equation}
where fitting formula for the free parameters $f_t$, $r_{te}$ and $\delta$ are as follows:
\begin{equation} \label{eq:green}
	\begin{aligned}
	f_t &= f_{\rm b}^{{a_1}^{\left(\frac{c}{10}\right)^{a_2}}} c^{a_3(1-f_{\rm b})^{a_4}} \\
	r_{te} &= r_{\rm vir}  f_{\rm b}^{{b_1}^{\left(\frac{c}{10}\right)^{b_2}}} c^{b_3(1-f_{\rm b})^{b_4}} \exp\left[ b_5 \left( \frac{c}{10}\right)^{b_6}(1-f_{\rm b})\right] \\
	\delta &= c_0  f_{\rm b}^{{c_1}^{\left(\frac{c}{10}\right)^{c_2}}} c^{c_3(1-f_{\rm b})^{c_4}} \,\,\,,
	\end{aligned}
\end{equation}
with the coefficients listed in Table~\ref{tab:greenparams}.

\begin{table}
	\caption{The best-fit parameters for Equation~\eqref{eq:green}, calibrated by \citet{green2019}.}
	\begin{center}
		\begin{tabular}{cc|cc|cc}
			\hline
			$a_1$  &  0.338  &  $b_1$  &  0.448  &  $c_0$  &  2.779  \\
			$a_2$  &  0.000  &  $b_2$  &  0.272  &  $c_1$  & -0.035  \\
			$a_3$  &  0.157  &  $b_3$  & -0.199  &  $c_2$  & -0.337  \\
			$a_4$  &  1.337  &  $b_4$  &  0.011  &  $c_3$  & -0.099  \\
			&         &  $b_5$  & -1.119  &  $c_4$  &  0.415  \\
			&         &  $b_6$  &  0.093  &         &         \\
			\hline
		\end{tabular}
\label{tab:greenparams}
\end{center}
\end{table}

\section{Effects of Shape on Stripping} \label{sec:shape}

\begin{figure*}
	\centering
	\subfloat{{\includegraphics{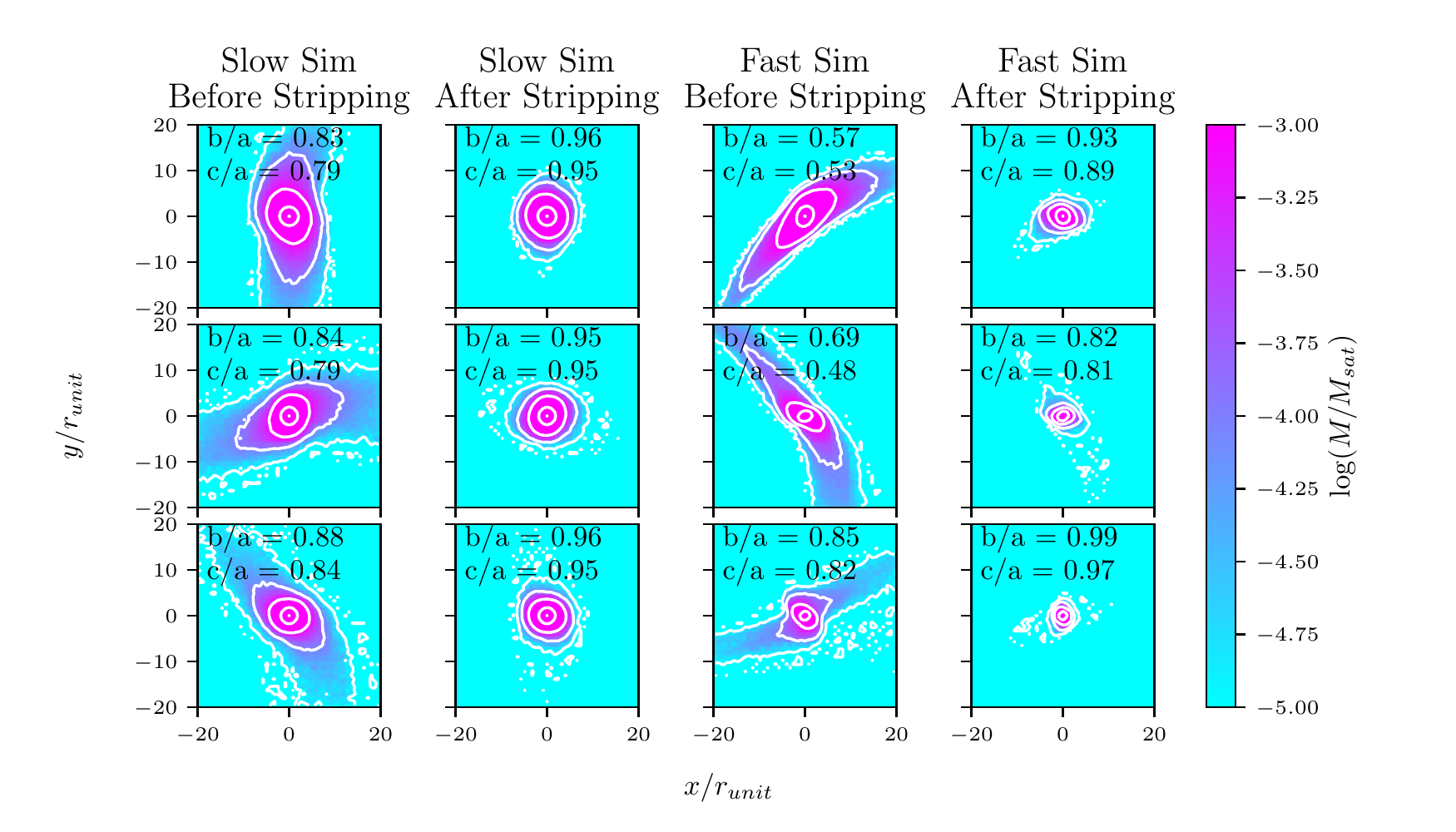}}}%
	
	\subfloat{{\includegraphics{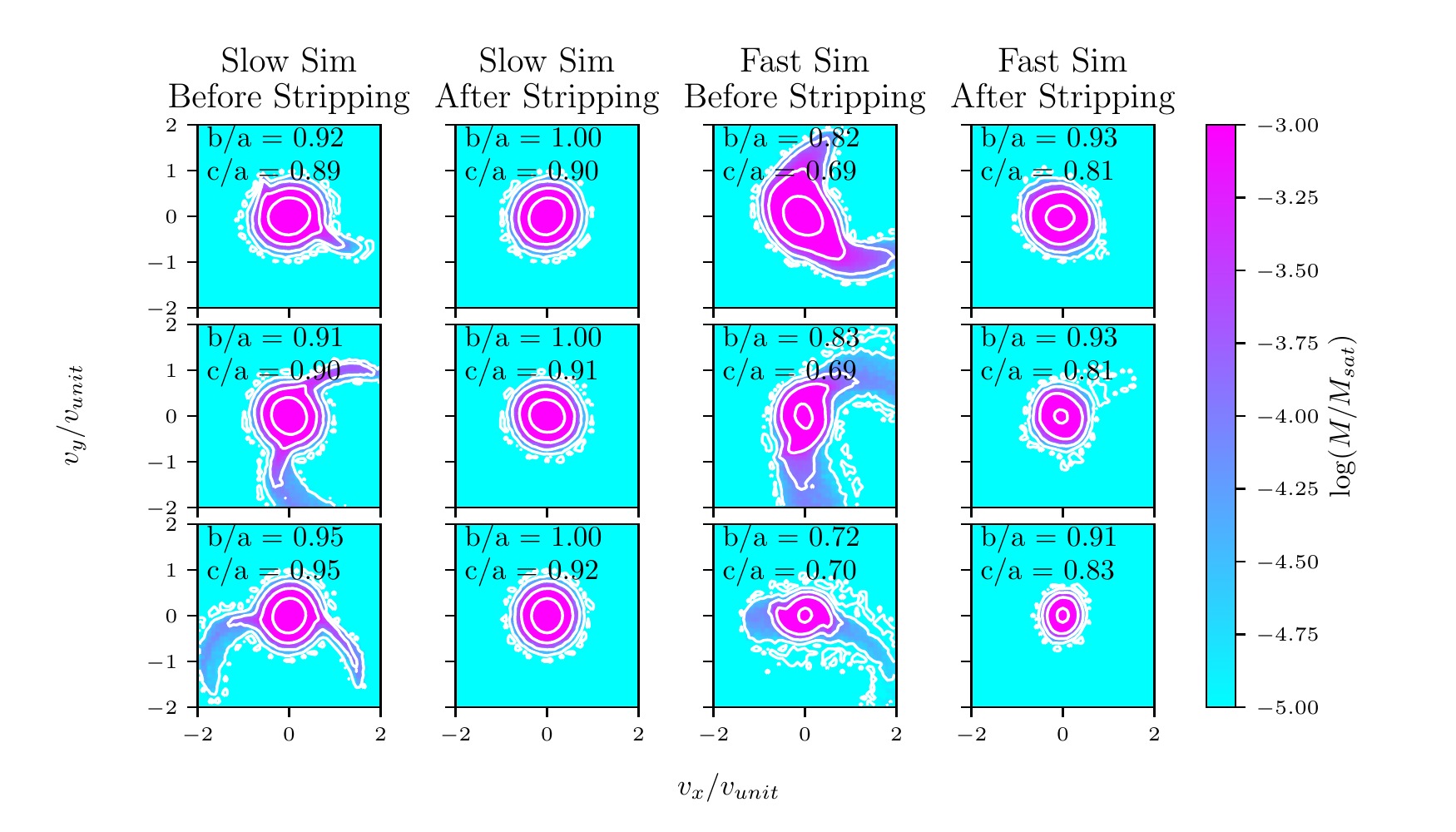}}}%
	\caption{ The subhalo at the first three pericentric passages (top to bottom) for both the Slow and Fast Simulations, and the corresponding shape ratios $b/a$ and $c/a$. {\color{black}The total mass in each $x$--$y$ bin is indicated as coloured, and contours of equal mass are shown in white.} Shape measurements in position space (top) and velocity space (bottom) are either for all particles bound at the previous apocentre (Before Stripping) or for those still bound at the subsequent apocentre (After Stripping).}
	\label{fig:Shape_remove}
\end{figure*}

In Section~\ref{sec:equil}, we showed that the satellite remnant becomes less spherical as it passes through pericentre; this could indicate a scenario in which particles at larger radii are more likely to be stripped, or alternatively, it could be that the sub-halo becomes more spherical as it settles into equilibrium around apocentre. We examine this in Fig.~\ref{fig:Shape_remove}, where we show the remnant as it passes through pericentre on the first three orbits (rows), considering either the particles that were bound at the previous apocentric passage (`Before Stripping') or only the particles that will remain bound by the next apocentric passage (`After Stripping'). The particles that remain bound are very spherical, even at pericentre. This suggests that it is \emph{not} the case that the remnant becomes spherical as it reaches equilibrium, but rather tidal stripping preferentially removes particles that are further away, resulting in a more spherical remnant. However, by the time the remnant reaches pericentre, the particles that will be removed are already in a distorted shape; likely many of these particles are those that are less energetically bound and therefore more likely to be stripped. Similarly,  Fig.~\ref{fig:Shape_remove} shows this phenomenon in velocity space. Again, the particles that remain bound are more `spherical', indicating that stripping preferentially removes particles that are non-spherical in velocity space.

	\bsp	
	\label{lastpage}
\end{document}